\documentclass[preprint,12pt]{elsarticle}
\journal{}
\usepackage[utf8]{inputenc}
\usepackage[T1]{fontenc}
\PassOptionsToPackage{hyphens}{url}
\usepackage{longtable,booktabs,array,calc,soul,listings,xcolor,graphicx,amsmath,amssymb,url}
\usepackage{hyperref}
\usepackage{scalefnt}
\makeatletter
\g@addto@macro\UrlBreaks{\do\a\do\b\do\c\do\d\do\e\do\f\do\g\do\h\do\i\do\j\do\k\do\l\do\m\do\n\do\o\do\p\do\q\do\r\do\s\do\t\do\u\do\v\do\w\do\x\do\y\do\z\do\A\do\B\do\C\do\D\do\E\do\F\do\G\do\H\do\I\do\J\do\K\do\L\do\M\do\N\do\O\do\P\do\Q\do\R\do\S\do\T\do\U\do\V\do\W\do\X\do\Y\do\Z\do\0\do\1\do\2\do\3\do\4\do\5\do\6\do\7\do\8\do\9}
\makeatother
\providecommand{\tightlist}{\setlength{\itemsep}{0pt}\setlength{\parskip}{0pt}}
\lstdefinelanguage{Solidity}{
  keywords={pragma,solidity,contract,function,public,private,internal,external,view,pure,payable,returns,return,if,else,for,while,mapping,address,uint,uint256,int,bool,bytes,bytes32,string,memory,storage,calldata,require,revert,emit,event,modifier,constructor,override,virtual,interface,library,using,is,new,true,false,msg,block,tx},
  morecomment=[l]{//},
  morecomment=[s]{/*}{*/},
  morestring=[b]"
}
\begin{document}

\begin{frontmatter}

\title{SuperPaymaster: Eliminating Centralized Signer Authority via Asset-Oriented Abstraction to Reconcile Usability and Decentralization in Account Abstraction}

\author[icdi]{Huifeng Jiao}
\ead{huifeng\_jiao@cmu.ac.th}

\author[icdi]{Nathapon Udomlertsakul\corref{cor1}}
\ead{nathapon.u@icdi.cmu.ac.th}
\cortext[cor1]{Corresponding author.}

\affiliation[icdi]{organization={International College of Digital Innovation (ICDI), Chiang Mai University},
  city={Chiang Mai},
  country={Thailand}}

\begin{abstract}
Most production ERC-4337 Paymasters rely on Process-Oriented Abstraction (POA): a centralized off-chain server signs each sponsorship request, acting as a potential censorship bottleneck. We propose Asset-Oriented Abstraction (AOA), encapsulating payment capability in a persistent, user-owned on-chain asset---the Gas Card---rather than an off-chain signing process. Following the Design Science Research (DSR) methodology, we implement SuperPaymaster on Optimism Mainnet, anchoring sponsorship validity in on-chain Soulbound Token state and deterministic policy rules, removing the off-chain signer as a validity gate. We evaluate gas costs via single-UserOp ERC-20 transfers on Optimism Mainnet (n = 50 per system). In pure L2 execution gas (\texttt{txGasUsed}; $\texttt{actualGasUsed} = \texttt{txGasUsed} + \texttt{PVG}$), SuperPaymaster (167,830) is lower than both evaluated POA baselines: Alchemy Gas Manager (205,951) and Pimlico ERC-20 paymaster (328,937). It still pays a \textasciitilde{}32,000-gas on-chain verification overhead versus Alchemy, but reduces gas by 49\% versus Pimlico by replacing on-chain token liquidation with an internal balance update. In total billed gas, SuperPaymaster (286,818) exceeds Alchemy (257,299) due to higher bundler PVG overhead, not paymaster architecture. Code structural analysis and on-chain Mainnet evidence confirm that sponsorship validity requires no off-chain signing server: \texttt{validatePaymasterUserOp} reads only on-chain state. These findings suggest that AOA can mitigate the usability--decentralization--efficiency trade-offs in gas payment.
\end{abstract}


\end{frontmatter}

\section{Introduction}

\subsection{Background and Problem Statement}

Blockchain technology, and Ethereum in particular \cite{Wood2014, Buterin2013Whitepaper}, introduced a model of decentralized and trust-minimized computation. To mitigate spam and compensate validators, Ethereum charges gas for every state-changing operation \cite{Wood2014, Nakamoto2008}. Gas is therefore a core primitive of Ethereum's security and economic design. However, the user-facing gas payment model remains a major usability bottleneck. Before completing a single application interaction, users may need to acquire ETH, often through a centralized exchange; navigate volatile fee markets; approve token spending allowances; and, in many cases, bridge assets across chains. Industry surveys and Ethereum roadmap analyses consistently identify gas fees as one of the top three barriers to dApp adoption, with abandonment rates exceeding 50\% when users encounter unexpected fee requirements \cite{Etherspot2025, Kim2024}. Early meta-transaction standards such as EIP-2771 \cite{Sandford2020} attempted to abstract gas payment at the application layer through trusted forwarder contracts, but they remained app-specific and lacked a unified sponsorship interface. ERC-4337 \cite{Buterin2021ERC4337} later introduced the Paymaster primitive to abstract gas payment at the pseudo-protocol layer, yet practical deployments remain constrained by architectural choices that reintroduce centralized dependencies.

\begin{figure}[htbp]
  \centering
  \includegraphics[width=0.85\textwidth]{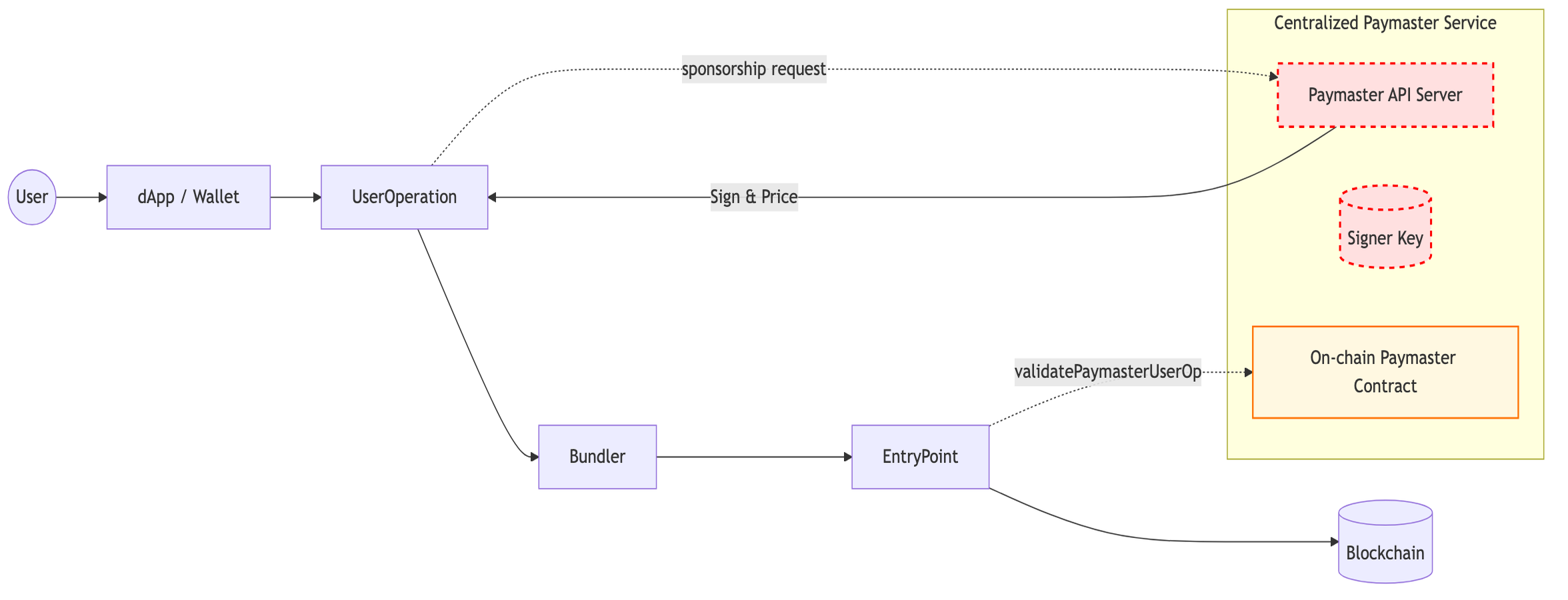}
  \caption{Standard ERC-4337 Paymaster Architecture. The centralized API server and signer (red, dashed) form the censorship bottleneck for gas sponsorship.}
  \label{fig:fig1_1_centralized_paymaster_architecture}
\end{figure}

Existing solutions exhibit two recurring limitations. First, most production Paymaster services rely on a centralized off-chain API server to sign each sponsorship request. If the server goes offline or blocks a specific user, the user is prevented from transacting, reintroducing a censorship vector at the sponsorship layer (see Figure~\ref{fig:fig1_1_centralized_paymaster_architecture}). BundleBear data \cite{BundleBear2024} shows that a single paymaster provider processes over 71\% of all sponsored UserOperations by operation count, while the top three providers account for over 62\% of total gas spend---concentrating sponsorship authority in a small number of API-based intermediaries. Second, existing abstractions are tied to a particular dApp, session, or token list, so a seamless experience in one application does not carry over to another. We term this dominant pattern \textbf{Process-Oriented Abstraction (POA)}: the system abstracts a discrete payment \emph{process} within a specific, limited context, rather than encapsulating a reusable payment \emph{capability}.

\begin{figure}[htbp]
  \centering
  \includegraphics[width=0.85\textwidth]{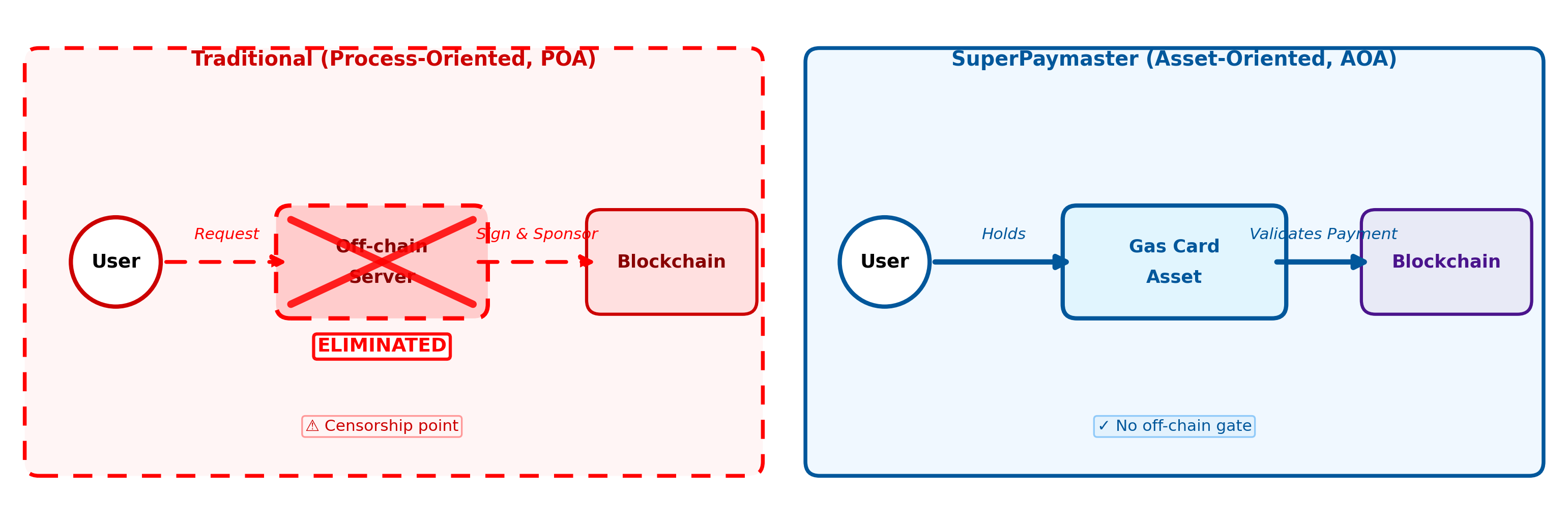}
  \caption{Conceptual Comparison. Traditional POA (red) requires an off-chain server to sign every transaction; SuperPaymaster (blue) validates sponsorship purely on-chain based on the user's Gas Card.}
  \label{fig:fig1_2_paradigm_comparison}
\end{figure}

A well-designed solution should reconcile an \textbf{intuitive user experience} (analogous to a prepaid card) with \textbf{decentralized on-chain verification}. This paper introduces \textbf{Asset-Oriented Abstraction (AOA)} as that reconciliation. By encapsulating payment permission within an on-chain asset rather than an off-chain signing process, AOA relocates authorization authority to the user's wallet. We operationalize the paradigm through \textbf{SuperPaymaster}, which instantiates the "Gas Card" metaphor according to three principles (Figure~\ref{fig:fig1_2_paradigm_comparison} contrasts the two architectures):

\begin{itemize}\tightlist
\item \textbf{Assetization}: sponsorship capability is encapsulated in a tangible digital asset---a Soulbound Token (SBT) \cite{Weyl2022DeSoc}, a non-transferable ERC-721. The Gas Card SBT extends the Soulbound Token concept into a gas-payment context, binding sponsorship eligibility to the user\textquotesingle{}s Smart Account rather than to any off-chain identity provider.
\item \textbf{Decentralization via Ownership}: because the SBT is an immutable, on-chain proof of payment capability, discretionary off-chain signatures from a Paymaster API are no longer required for sponsorship validity. Standard Bundlers still handle transaction submission, but the \emph{permission to pay} is enforced fully on-chain.
\item \textbf{Universal Mental Model}: holding a valid card enables portable payment across compatible dApps, without exposing relayer or bundler mechanics to the user.
\end{itemize}

\textbf{AOA does not replace POA at the protocol layer}; rather, it repositions what is abstracted. POA abstracts the \emph{process} (and exposes its steps to the user whenever the abstraction leaks), whereas AOA encapsulates the \emph{capability} in an asset the user owns. SuperPaymaster therefore still uses ERC-4337 as the execution pipeline; the contribution lies in the architectural shift in \emph{how eligibility is proven and where sponsorship authority resides}.

\subsection{Research Questions and Contributions}

This research aims to design, implement, and evaluate SuperPaymaster to answer:

\begin{itemize}\tightlist
\item \textbf{RQ1}: How does AOA eliminate the off-chain paymaster signer as a mandatory authorization checkpoint while maintaining a seamless user experience?
\item \textbf{RQ2}: How can the Gas Card metaphor reduce cognitive load and interaction complexity relative to POA approaches?
\item \textbf{RQ3}: To what extent does SuperPaymaster improve economic efficiency compared to EOA baselines and standard ERC-4337 Paymaster implementations?
\end{itemize}

This paper makes three contributions to blockchain HCI and system design:

\begin{enumerate}\tightlist
\item \textbf{Conceptual}. Definition and conceptualization of \textbf{Asset-Oriented Abstraction} as a design principle that systematically addresses the usability--decentralization--efficiency trilemma in gas payment mechanisms.
\item \textbf{Artifact}. The open-source \textbf{SuperPaymaster} system, comprising on-chain contracts (including the OpenCards SBT-based Gas Card standard and the OpenPNTs gas-token standard enabling permissionless community issuance) and a supporting SDK for UI integration. The artifact realizes AOA in two deployment modes: \textbf{PaymasterV4}, a fully self-deployable single-community paymaster that any community can run independently---issuing its own gas token and managing its own deposit, with no shared-infrastructure dependency and no off-chain signer---and \textbf{SuperPaymaster}, a multi-community shared public-infrastructure layer built on the same AOA core (detailed in §4.1, compared in Appendix~F); PaymasterV4 also serves as the internal cost-decomposition baseline (T1) in §5.3.
\item \textbf{Evaluation}. A multi-method evaluation on Optimism Mainnet (n = 50 per system, single-UserOp ERC-20 transfer) covering decentralization (code structural analysis of \texttt{validatePaymasterUserOp} and on-chain Mainnet tx evidence), usability (GOMS analytical modeling), and efficiency (gas profiling against industry POA baselines---Alchemy Gas Manager and Pimlico ERC-20 paymaster---with trace-level decomposition, bootstrap 95\% confidence intervals (CIs), and Cliff's $\delta$).
\end{enumerate}

All system components---contracts, SDK, and deployment scripts---are released as open-source infrastructure to support reproducibility and community adoption.

\subsection{Paper Structure}

The remainder of this paper is organized as follows:
\begin{itemize}\tightlist
\item \textbf{Section 2: Related Work} reviews the theoretical and conceptual foundations of HCI as applied to Web3 systems and analyzes the limitations of existing Process-Oriented solutions.
\item \textbf{Section 3: Methodology} outlines the Design Science Research (DSR) framework used to guide the creation and evaluation of the artifact.
\item \textbf{Section 4: Artifact Design} details the system architecture of SuperPaymaster, including the "Gas Card" (SBT) and the Zero-Approve Gas Token Architecture.
\item \textbf{Section 5: Evaluation} presents the empirical results from our mainnet experiments, focusing on gas efficiency, industry baselines, and paymaster-signer decentralization.
\item \textbf{Section 6: Discussion} interprets the findings through the lens of Asset-Oriented Abstraction and discusses the theoretical implications for Web3 UX.
\item \textbf{Section 7: Conclusion} summarizes the research contributions and outlines future directions.
\end{itemize}

\section{Related Work and Problem Analysis}

\subsection{Theoretical Foundations}

\subsubsection{An HCI Lens on Gas Payment}

The usability of blockchain systems is a well-documented challenge \cite{Frohlich2022, Saldivar2023, Alqaryouti2025}, and recent work in \emph{Blockchain: Research and Applications} emphasizes the need for accessible smart contract architectures that reduce operational friction without weakening security invariants. Norman's "gulf of execution" \cite{Norman2013, Vermeulen2013} captures the general gap between user intent and required actions; Nielsen's usability heuristics \cite{Nielsen1994} pinpoint specific violations in the typical gas workflow. Users encounter (i) a poor \emph{match between system and the real world}---"Gas Limit," "Gwei," and "Nonce" have no natural payment analog, whereas an auction-based fee market has no analog in users' everyday payment experience; (ii) low \emph{flexibility and efficiency of use}---a single intent is fragmented across Approve/Swap/Bridge/Execute steps; and (iii) inadequate \emph{error prevention}---users routinely fall into "insufficient ETH for gas" dead ends even when holding sufficient token balances. These are systemic HCI failures, not local rough edges. A principled solution must therefore redesign the payment surface, not patch individual steps.

\subsubsection{Perceived Ease of Use as a Gatekeeper}

The Technology Acceptance Model \cite{Davis1989, Marangunic2015} holds that adoption is driven by Perceived Usefulness (PU) and Perceived Ease of Use (PEOU). In Web3, PEOU functions as a gatekeeper to PU: users cannot appreciate a DeFi or gaming dApp's utility if they cannot successfully execute a first transaction. The recurring friction of gas management therefore blocks users from ever reaching the point of value discovery, which is a primary driver of low retention and high drop-off in dApp onboarding funnels.

\subsubsection{From Process-Oriented to Asset-Oriented Abstraction}

The core HCI defect in current gas solutions is the prevalence of \emph{leaky abstractions}. We categorize existing solutions under a paradigm we term \textbf{Process-Oriented Abstraction (POA)}: the system abstracts a discrete payment \emph{process} within a specific, limited context. Because the abstraction is tied to a dApp, session, or token list, whenever the scoped context is exceeded, the underlying steps re-surface to the user. In contrast, we propose \textbf{Asset-Oriented Abstraction (AOA)}: the system transforms the recurring payment process into a persistent, user-owned digital asset---the Gas Card. AOA is defined by three core principles (§1.1), which at the implementation layer decompose into three technical requirements: Assetization (the capability is reified as a standalone contract), User-Ownership (the asset's state is controlled by the user's Smart Account rather than a service contract), and Deterministic Verification (eligibility is evaluated against on-chain state under ERC-4337 validation constraints, without off-chain signatures).

Two clarifications matter here. First, AOA is not a rejection of ERC-4337; SuperPaymaster is a concrete ERC-4337 Paymaster. AOA reframes what the abstraction \emph{targets}---capability rather than process---and therefore changes where sponsorship authority resides. Second, POA and AOA describe design stances; they coexist on-chain today, and AOA is compatible with future Native AA proposals.

Positioning AOA on the decentralization spectrum is useful here, because decentralization is not a binary state but a gradient along multiple dimensions (architectural, political, logical) \cite{Buterin2017Decentralization, Walch2019}. Current Paymaster deployments are architecturally decentralized (contract logic on-chain) but politically centralized (an off-chain signer decides \emph{who} gets sponsored). AOA moves the political axis on-chain as well: the service logic that decides sponsorship eligibility is fully encapsulated in contracts, which aligns with the "hyperstructure" or \emph{code-as-law} end-state described by Jacobs \cite{Jacobs2022}, where service logic is permissionless, immutable, and self-sustaining.

\subsubsection{The Bootstrapping Paradox: A Preemptive Defense}

A natural critique of asset-based abstractions is the \emph{Bootstrapping Paradox}: does requiring a Gas Card merely shift the friction of acquiring ETH onto acquiring the Card? We argue that the acquisition channel is categorically different. Acquiring ETH almost always involves Centralized Exchanges and fiat on-ramps---"Centralized Gatekeepers" constrained by KYC, geography, and banking infrastructure. In the AOA paradigm, a Gas Card (SBT) is permissionlessly self-minted by any user directly from the MySBT protocol contract---no community authorization is required, only a chain account and a small GToken balance. Community gas tokens (xPNTs) are separately issued by each participating community using open-source infrastructure; any DAO, game, or wallet can deploy xPNTs and a paymaster operator permissionlessly, and users holding a Gas Card can access sponsorship from any such community. Card acquisition is \textbf{one-time rather than per-transaction}: once a user holds a Card, subsequent sponsored transactions proceed without any further acquisition step, converting an otherwise perpetual per-transaction friction into an amortized onboarding cost. We therefore do not eliminate the initial "acquire an asset" action; instead, we move it from a recurring, gatekeeper-mediated cost to a one-time, permissionless setup. This shift is structurally decentralized, not a superficial relabeling.

\subsection{Technical Foundations: Account Abstraction and Fee Mechanisms}

\begin{figure}[htbp]
  \centering
  \includegraphics[width=0.85\textwidth]{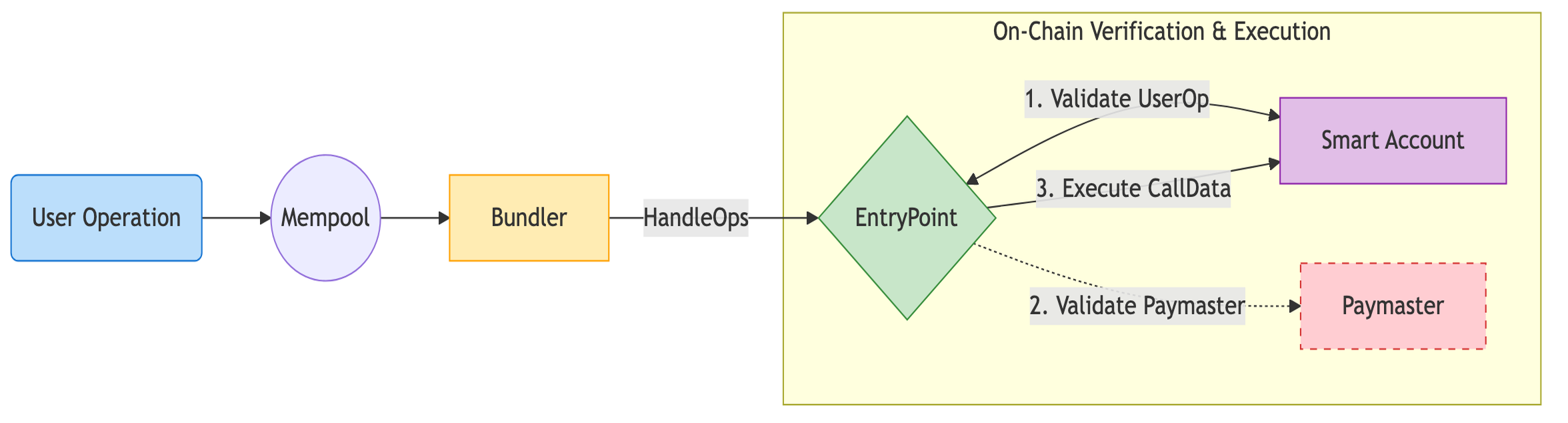}
  \caption{Standard ERC-4337 Architecture. The EntryPoint coordinates validation between the Smart Account and optional Paymaster.}
  \label{fig:fig2_1_erc4337_architecture}
\end{figure}

Ethereum wallet addresses have reached 300 million \cite{Etherscan2024}. Account Abstraction (AA), and ERC-4337 in particular \cite{Buterin2021ERC4337}, introduces the Paymaster primitive for gas sponsorship. The ERC-4337 pipeline routes a \texttt{UserOperation} through a mempool and a Bundler into the \texttt{EntryPoint}, which validates the Smart Account and the optional Paymaster before execution (Figure~\ref{fig:fig2_1_erc4337_architecture}); reference implementations of the singleton entrypoint and verifying paymaster are available \cite{Singh2023, Tirosh2022}.

\subsubsection{Account Abstraction Standards in Context}

\begin{sloppypar}
The AA ecosystem now spans three tiers. Application-layer standards include ERC-4337 (the dominant production standard) and ERC-7562 (validation-phase restrictions protecting bundlers from DoS). Bridge mechanisms include EIP-7702 \cite{eip7702} in the Prague/Electra (Pectra) upgrade (activated May 7, 2025), which allows existing EOAs to temporarily function as smart accounts \cite{Buterin2024Wallet}, alongside Passkeys/WebAuthn and Wallet-as-a-Service \cite{a16z2024, ZeroDev2023Sessions, ParticleNetwork2023WaaS}. Emerging protocol-level proposals for Native AA include RIP-7560 \cite{rip7560} with supporting specifications RIP-7711 (bundle transaction type) and RIP-7712 (two-dimensional nonce) for L2s, EIP-7701 \cite{eip7701} for Ethereum Mainnet, and EIP-8141 Frame Transaction \cite{EIP8141_2025}, a Draft proposal under discussion for the Glamsterdam upgrade as of early 2026. While Native AA promises greater efficiency, ERC-4337 remains the dominant production standard today. Our research optimizes the user experience within ERC-4337 while remaining forward-compatible with Native AA. Surveys of ERC-4337 \cite{Wang2023, Lin2024} report that its security benefits (social recovery, multi-sig validation) are widely recognized, but its \emph{usability} impact remains mixed: Paymasters are typically deployed as Service Providers rather than Asset Issuers, producing session-scoped sponsorship that fails to compose across dApps \cite{OpenZeppelin2024}.
\end{sloppypar}

Native AA proposals absorb the bundler role into the execution layer but remain silent on who decides sponsorship eligibility. The paymaster-signer authority question---which user gets sponsored and under what conditions---is orthogonal to bundler-layer changes and persists regardless of the native AA timeline. AOA therefore retains independent contribution value in a post-Native-AA world.

\subsubsection{Transaction Fee Mechanisms and the Mental Model Gap}

\begin{sloppypar}
The transition from first-price auctions to EIP-1559 improved the \emph{predictability} of inclusion \cite{Roughgarden2024} but introduced a two-variable cognitive burden (base fee + tip) that empirical analyses find non-trivial for non-expert users \cite{Liu2022EIP1559}. At a more fundamental level, the mental model of "bidding for blockspace" is misaligned with users' everyday expectation of "paying a fixed price" or using a prepaid service \cite{Saldivar2023}; this mismatch is a direct source of cognitive friction, as described by Sweller \cite{Sweller1988}. This motivates an asset-based mental model in which the user sees balance rather than bids.
\end{sloppypar}

\subsection{State-of-the-Art Analysis}

\subsubsection{Incomplete Abstractions in Current Solutions}

Existing academic and industry solutions from providers such as Pimlico \cite{Pimlico2023}, Alchemy \cite{Alchemy2023}, Biconomy \cite{Biconomy2023}, Stackup \cite{Stackup2023}, and Coinbase \cite{Coinbase2023} have driven real progress in modular smart accounts \cite{Safe2024}, but most present an incomplete abstraction characterized by two recurring properties. First, \emph{leaky abstraction}: users may no longer need ETH, but may need to acquire specific dApp-supported stablecoins or manage per-application gas tanks. Second, \emph{fragmented experience}: gas sponsorship is typically tied to a specific dApp or vendor, so a seamless experience in one application does not carry over to another.

\subsubsection{Industry Implementations and Market Concentration}

Table 1 compares existing solutions against SuperPaymaster along five dimensions that matter for the research questions.

\begin{table}[htbp]\centering
\resizebox{\textwidth}{!}{%
\begin{tabular}{|l|l|l|l|l|}
\hline
\textbf{Dimension} & \textbf{Alchemy \cite{Alchemy2023}} & \textbf{Pimlico \cite{Pimlico2023}} & \textbf{Biconomy \cite{Biconomy2023}} & \textbf{SuperPaymaster (Ours)} \\
\hline
Core Paradigm & Process-Oriented & Process-Oriented & Process-Oriented & \textbf{Asset-Oriented} \\
Sponsorship Gate & Off-chain API signature & Off-chain API signature & Off-chain API signature & \textbf{On-chain SBT + deterministic policy} \\
Custom ERC-20 Gas & Yes (on-chain swap) & Yes (on-chain swap) & Limited (restricted token list) & \textbf{Yes (Zero-Approve, no swap)} \\
Paymaster-Signer Censorship Removed & No & No & No & \textbf{Yes} \\
Community-Issued & Limited & No & Limited & \textbf{Yes (Permissionless)} \\
\hline
\end{tabular}
}
\end{table}

\noindent\textit{Table 1: Comparison of Industry Account Abstraction Solutions.}

\begin{figure}[htbp]
  \centering
  \includegraphics[width=0.64\textwidth]{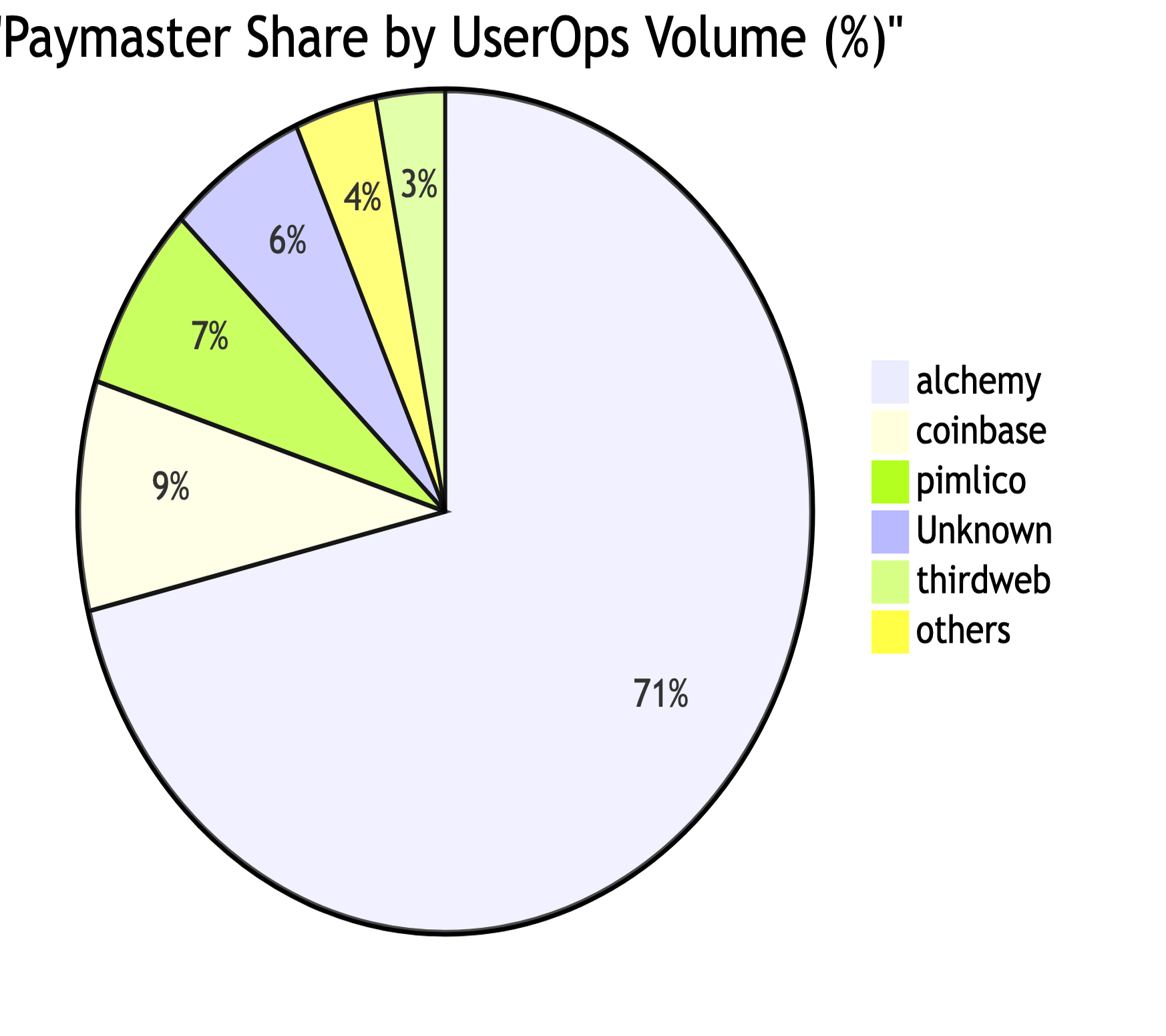}
  \caption{Paymaster market concentration by UserOps volume. (Data source: BundleBear ERC-4337 Paymasters \cite{BundleBear2024}, retrieved February 2026.)}
  \label{fig:fig2_2_paymaster_market_userops}
\end{figure}

\begin{figure}[htbp]
  \centering
  \includegraphics[width=0.60\textwidth]{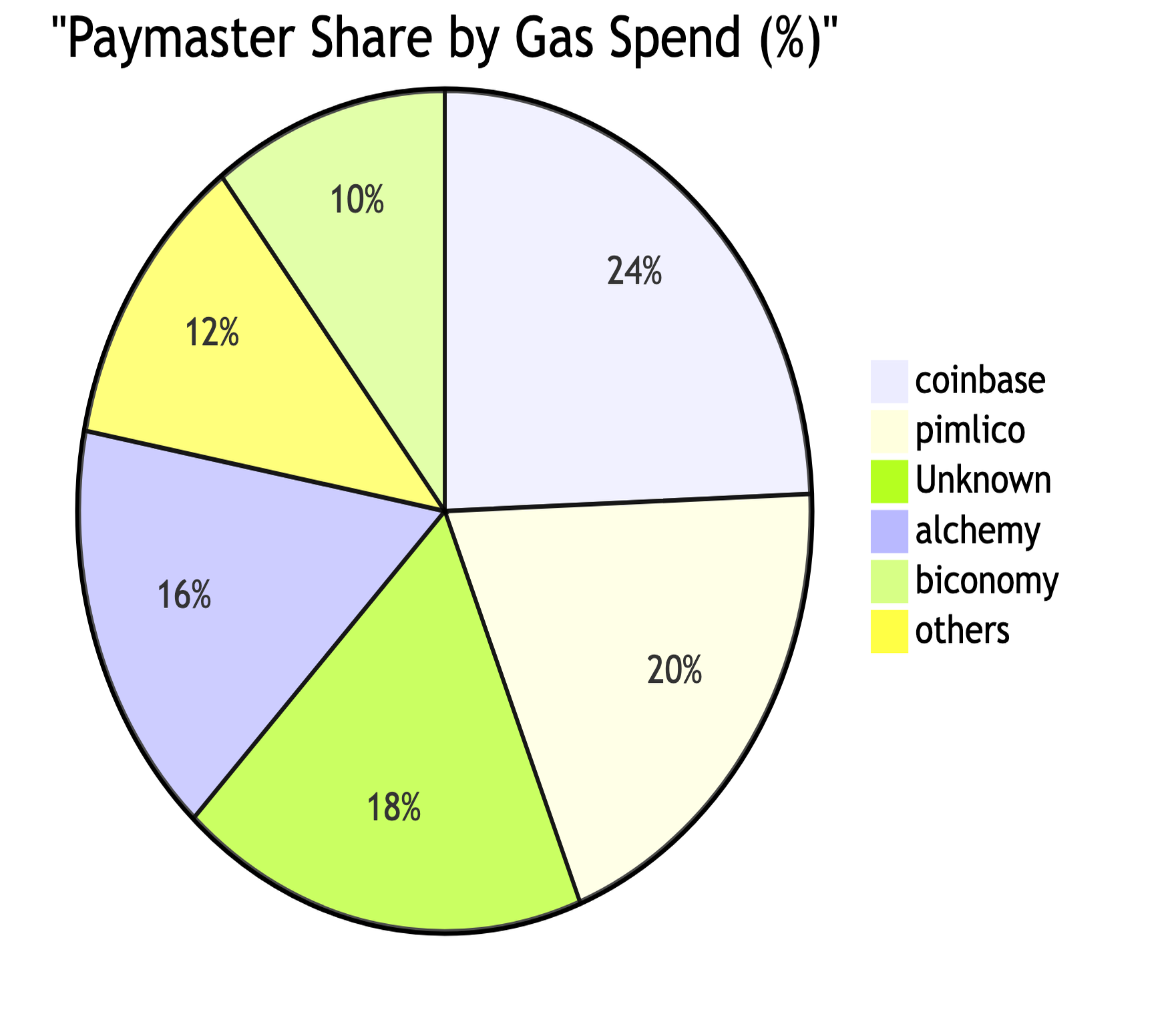}
  \caption{Paymaster market concentration by gas spend. (Data source: BundleBear~\cite{BundleBear2024}, retrieved February 2026.)}
  \label{fig:market-gas-spend}
\end{figure}

The ERC-4337 paymaster market shows significant concentration across both dimensions. By transaction volume (Figure~\ref{fig:fig2_2_paymaster_market_userops}), a single provider (Alchemy) processes over 71\% of all sponsored UserOperations---largely driven by high-volume, low-cost batch operations. By economic weight (total gas spend, Figure~\ref{fig:market-gas-spend}), the distribution is more balanced but still concentrated: the top three providers (Coinbase, Pimlico, Unknown) collectively account for over 62\% of all paymaster gas expenditure \cite{BundleBear2024}. Both dimensions create long-term risks of monopoly, censorship, and systemic vulnerabilities analogous to MEV centralization \cite{Daian2020}.

\subsubsection{Cost and Scalability Considerations}

High AA gas costs on Layer 1 necessitate Layer 2 deployment. Thibault et al. \cite{Thibault2022} show that rollups reduce fees by 20--100x relative to Ethereum Mainnet, so we target L2 and report efficiency primarily in L2 execution gas units, complemented by receipt-level fee decomposition on OP Stack (L2 execution + L1 data availability + protocol overhead) when discussing end-user cost.

\subsection{Research Gap}

A clear gap emerges from this review: \textbf{to the best of our knowledge, no existing gas payment solution simultaneously (i) reduces cognitive load (improving PEOU) and (ii) removes off-chain paymaster-signer authority from the sponsorship validity path.} POA solutions trade trust-boundary size for convenience through centralized intermediaries; EOA models preserve decentralization but fail on usability. SuperPaymaster targets this gap through Asset-Oriented Abstraction instantiated as a Gas Card, offering prepaid-card usability with deterministic on-chain eligibility rules.

\section{Methodology}

This research follows the \textbf{Design Science Research (DSR)} methodology of Peffers et al. \cite{Peffers2007}, anchored by the seven design-science guidelines of Hevner et al. \cite{hevner2004design} and the Framework for Evaluation in Design Science (FEDS) of Venable et al. \cite{venable2016feds}. Per FEDS, our evaluation is \textbf{ex-post} (evaluating a deployed artifact), \textbf{artificial} (controlled test harnesses and selected mainnet UserOps rather than broad in-the-wild usage), and \textbf{summative} (assessing whether the artifact achieves its design objectives). Our process iterates through the six standard DSR activities: problem identification (§1), objective definition (§2), design and development (§4), demonstration (controlled local environment plus Optimism Mainnet, §5), evaluation (mixed-evidence measurement against baselines, §5), and communication (§6--7).

\subsection{Evaluation Strategy}

To answer the research questions, we employ \textbf{Empirical Benchmarking and Analytical Evaluation}. Empirical Gas Profiling executes real transactions on Optimism Mainnet and a local Anvil network to produce deterministic, receipt-level gas measurements. Economic Scenario Analysis applies real-world market variables (gas-price and ETH-price volatility) to those measurements to reason about cost sensitivity under varying network conditions.

\subsubsection{Experimental Setup}

We define three workflows for a standard ERC-20 "Token Transfer" task:

\begin{itemize}\tightlist
\item \textbf{Baseline A}: Standard EOA ERC-20 Transfer on Optimism Mainnet (qualitative grounding; not directly comparable to UserOperations because execution paths differ fundamentally).
\item \textbf{Baseline B}: Centralized gasless services using standard ERC-4337 paymasters (Alchemy as API-verified representative; Pimlico as DEX-routed ERC-20 paymaster representative), sampled from on-chain \texttt{User\-Operation\-Event} logs filtered for single-UserOp bundles with ERC-20 \texttt{transfer} selector.
\item \textbf{Workflow C (SuperPaymaster)}: The user holds a Gas Card (SBT) and xPNTs gas tokens and simply initiates the transaction.
\end{itemize}

For Workflow C we executed repeated transactions on Optimism Mainnet and logged receipt-level fee decomposition for two operation types. \textbf{T1} is a gasless token transfer via PaymasterV4 in prepayment mode; \textbf{T2.1} is a gasless payment via SuperPaymaster in normal mode, which burns the user's community gas token.\footnote{The operation-type labels follow our \texttt{aastar-sdk} dataset convention, where \textbf{T2} denotes the credit mode and \textbf{T2.1} the normal mode of SuperPaymaster. This paper evaluates the normal mode (T2.1); the credit mode (T2) is studied in a companion paper. We retain the original label to keep this paper consistent with the archived dataset.} Each operation type was executed \textbf{N = 50} times to support descriptive statistics, bootstrap confidence intervals, and non-parametric effect size estimation in §5. The primary evaluation compares SuperPaymaster (T2.1) against industry POA baselines (Alchemy Gas Manager and Pimlico ERC-20 paymaster, n = 50 each) to validate AOA's economic viability relative to the dominant process-oriented paradigm. The T1 (PaymasterV4) data provides an internal cost-decomposition reference: both PaymasterV4 and SuperPaymaster belong to the same AOA family and both remove the off-chain signer dependency, so contrasting T1 and T2.1 isolates the gas overhead of SuperPaymaster's additional checks (SBT verification, rate limits, credit state).

\subsubsection{Evaluation Propositions and Metrics}

Rather than framing the study as formal statistical hypothesis testing, we evaluate three \emph{propositions} (P1--P3). Each proposition states a claim about the artifact and is validated by a corresponding evaluation in §5 (P1 in §5.1, P2 in §5.2, P3 in §5.3):

\begin{itemize}\tightlist
\item \textbf{P1 (Paymaster-signer gate removal).} A Gas Card-based sponsorship eligibility rule enables transactions to remain valid without a discretionary off-chain paymaster signing service; sponsorship validity is determined entirely by on-chain contract state, making the off-chain signer architecturally unreachable as a validity gate.
\item \textbf{P2 (Analytical usability improvement).} The modeled workflow reduces interaction steps and extraneous cognitive operators related to gas management (GOMS-based analytical evaluation).
\item \textbf{P3 (Comprehensive-cost trade-off).} Moving sponsorship authorization on-chain introduces measurable validation overhead relative to industry POA paymasters that rely on lightweight off-chain ECDSA signing, but can reduce comprehensive user cost by eliminating DEX-routed token liquidation paths and recurring ETH-balance management friction.
\end{itemize}

We measure three primary dependent variables: Interaction Steps (S), a proxy for cognitive load; Total Time (T) for efficiency; and Total Cost (C) including gas fees, exchange fees, and service premiums. Decentralization (P1) is assessed qualitatively via contract structural analysis (presence or absence of off-chain \texttt{ecrecover} in the validation path) rather than a binary trial metric.

\subsubsection{Data Analysis}

Gas measurements are deterministic on-chain values tied to a specific execution harness (contract version, chain state, and bundler configuration; see Appendix B), so we treat them as engineering measurements and do not apply inferential significance testing \cite{Jain1991}. We report descriptive statistics (mean, median, $\sigma$), bootstrap 95\% confidence intervals (10,000 resamples) \cite{Efron1993} to quantify estimation uncertainty without distributional assumptions, non-parametric \textbf{Cliff's $\delta$} with standard thresholds (|$\delta$| < 0.147 negligible, < 0.33 small, < 0.474 medium, $\geq$ 0.474 large) \cite{Romano2006} for pairwise effect sizes, and skewness/excess kurtosis for distribution diagnostics. Non-overlapping 95\% CIs between systems are treated as evidence of a robust difference, consistent with engineering measurement conventions \cite{Jain1991}.

\section{Artifact Design}

\subsection{Scope and Reproducibility}

The evaluated artifact comprises on-chain components (Registry, SuperPaymaster, PaymasterV4, Gas Card SBT, xPNTsFactory, and an optional BLS Aggregator for future decentralized governance) and off-chain components (ERC-4337 Bundler, community Relayer/Operator, and SDK data-collection scripts). The Relayer is not a validity gate: payment eligibility is enforced on-chain. Canonical deployment addresses are listed in Appendix~A, annotated core contract logic in Appendix~B, source repositories in the Data Availability statement, and exact reproduction commands in Appendix~C.

The artifact enforces four invariants that tie its behavior to its threat model (§4.7):

\begin{enumerate}\tightlist
\item \textbf{No discretionary signature.} Eligibility is determined by contract state (SBT ownership plus configured token policy), not by an operator signature that can censor specific users.
\item \textbf{Cost cap.} A cost cap bounds paymaster exposure per operation and uses the capped value for settlement.
\item \textbf{Token sufficiency.} A token sufficiency check prevents sponsorship unless the user can settle the required amount under supported-token and allowance constraints.
\item \textbf{Replay safety.} Replay safety at the AA layer is enforced by ERC-4337 UserOperation nonce rules, so the paymaster validates only sponsorship-specific constraints.
\end{enumerate}

\subsubsection{Two AOA Modes: PaymasterV4 and SuperPaymaster}

The artifact provides two deployment modes within the AOA paradigm. \textbf{PaymasterV4} is a single-community paymaster: the community deploys its own contract, manages its own ETH deposit, and configures a supported gas token. It requires Solidity development and DevOps capacity for deployment, funding, and monitoring. \textbf{SuperPaymaster} is a multi-community public infrastructure: communities register via a shared Registry and issue gas tokens through the xPNTsFactory without deploying any contracts. An operator manages the shared ETH deposit pool, reducing per-community operational overhead to near zero.

Both modes remove the off-chain paymaster signer and share the AOA core property of on-chain eligibility verification. SuperPaymaster extends PaymasterV4 with SBT-based identity binding, rate-limit governance, and a multi-community management interface---features that support permissionless multi-community use but add \textasciitilde{}13k gas to the validation step (§5.3). The trade-off is explicit: higher on-chain verification cost in exchange for lower operational cost and broader accessibility. Table~F.1 (Appendix~F) provides a detailed feature-by-feature comparison.

\subsection{Design Principles}

SuperPaymaster is built on four design principles, three grounded in HCI theory \cite{Shneiderman2010, Nielsen2013Personas, Hollender2010} and one in the ecosystem's structural requirement:

\begin{enumerate}\tightlist
\item \textbf{Asset-Centricity.} Encapsulate sponsorship capability in a persistent, portable asset (the Gas Card), shifting control from service provider to user.
\item \textbf{Metaphor Consistency.} Adhere strictly to the prepaid-card metaphor: a Card works across merchants in the physical world and, in the on-chain analogue, across dApps that integrate the SuperPaymaster sponsorship interface.
\item \textbf{Invisible Complexity.} Bridge Norman's gulf of execution \cite{Norman2013} by hiding the machinery---users see a balance, not Gwei or gas limits.
\item \textbf{Permissionless Open Infrastructure.} All core components are released under an open-source licence. Users permissionlessly self-mint the Gas Card SBT from the protocol-level MySBT contract; communities permissionlessly deploy their own paymaster operator and gas token (xPNTs, or wrap an existing ERC-20) via the open-source contracts, with no gatekeeper authority over either action; broad adoption across heterogeneous communities has not yet been empirically verified.
\end{enumerate}

\subsection{System Architecture}

\begin{figure}[htbp]
  \centering
  \includegraphics[width=\textwidth]{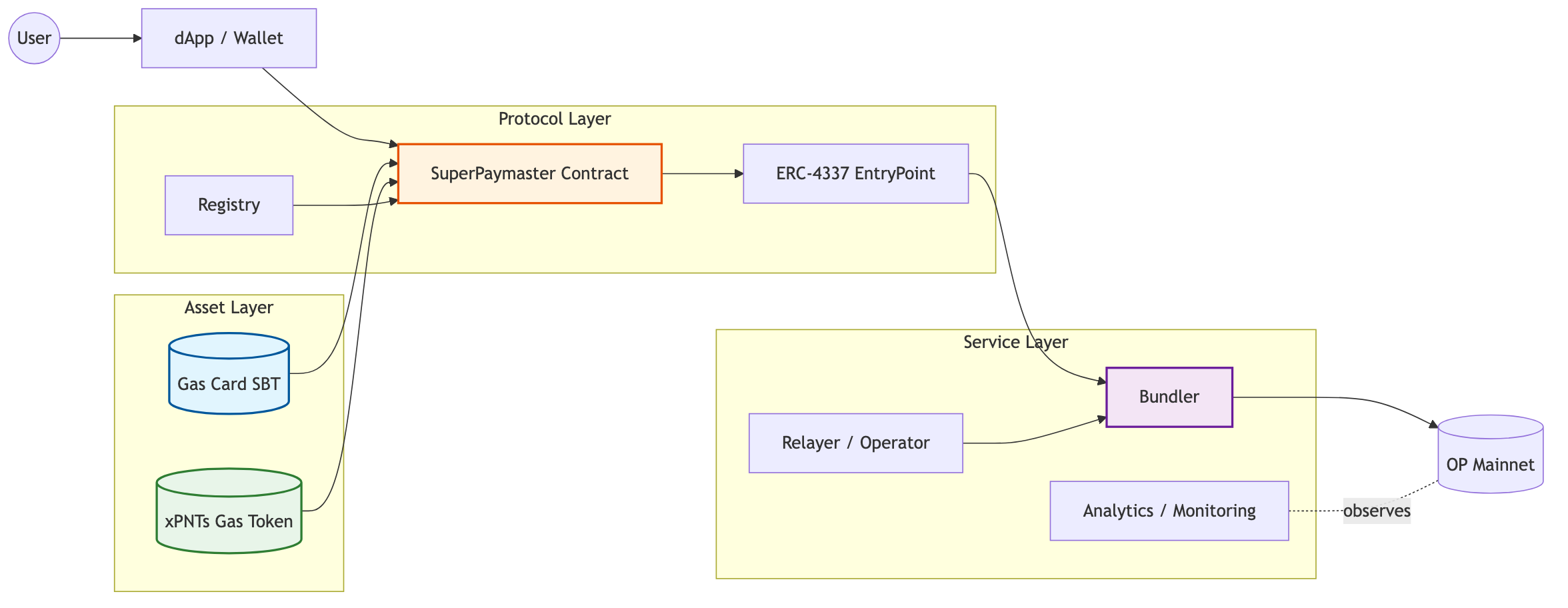}
  \caption{SuperPaymaster three-layer architecture. Sponsorship validity is enforced at the Protocol Layer; the Service Layer handles inclusion, not authorization.}
  \label{fig:fig4_1_system_architecture}
\end{figure}

The SuperPaymaster architecture is composed of three layers (Figure~\ref{fig:fig4_1_system_architecture}). The \textbf{Asset Layer} holds the Gas Card SBT (a non-transferable ERC-721 bound to a Smart Account, preventing theft and keeping reputation with the user) and the xPNTs gas token (an ERC-20 following the OpenPNTs protocol). The \textbf{Protocol Layer} contains the Registry, the SuperPaymaster contract (the on-chain verifier integrated with the ERC-4337 \texttt{EntryPoint}), and the \texttt{EntryPoint} itself. The \textbf{Service Layer} contains the Bundler, community Relayers/Operators, and Analytics/Monitoring services; the Service Layer is not a validity gate.

\subsubsection{Paymaster Validation Pipeline}

SuperPaymaster's sponsorship validity does not rely on a discretionary off-chain signing server. Instead, it is enforced by deterministic on-chain state evaluation under ERC-4337 validation constraints. The validation pipeline proceeds through five stages:

\begin{sloppypar}
\begin{enumerate}\tightlist
\item \textbf{Identity and asset verification.} Eligibility is checked against on-chain Gas Card ownership and/or cached eligibility state. If \texttt{SBT\_Balance(sender) == 0} or \texttt{eligibility[sender] == false}, the transaction reverts. This step reads up to three storage slots (SBT ownership, eligibility cache, rate-limit counter) and constitutes the primary gas overhead relative to a minimal paymaster that performs only ECDSA verification.
\item \textbf{Operator configuration resolution.} The paymaster resolves the operator's configured parameters---supported token list, exchange rate, per-card spending cap, and rate-limit window---from the Registry. These reads are O(1) SLOAD operations but contribute \textasciitilde{}5k gas cumulatively.
\item \textbf{Cost estimation and capping.} The paymaster computes \texttt{CappedCost = min(RequestedMaxCost, ProtocolHardCap)}, bounding exposure per operation.
\item \textbf{Settlement capability verification.} If \texttt{GasToken\_Balance(sender) < CappedCost}, the transaction reverts. This ensures the user can settle the expected cost through configured gas-token rules.
\item \textbf{Execution and accounting.} Settlement is applied via atomic token burn (xPNTs) and internal treasury transfer (aPNTs) under the configured policy mode; \texttt{EntryPoint} then executes the primary payload.
\end{enumerate}
\end{sloppypar}

Trace-level analysis reveals that \texttt{validatePaymasterUserOp} consumes approximately 48,625 gas for SuperPaymaster---compared to \textasciitilde{}16,000 gas for the off-chain ECDSA check in industry API-signature paymasters (e.g., Alchemy Gas Manager) and \textasciitilde{}35,549 gas for the same-family PaymasterV4. The \textasciitilde{}32k gas delta against ECDSA-based POA paymasters is the measurable on-chain execution cost of replacing the off-chain signer with deterministic state verification (SBT ownership, rate-limit counters, operator configuration, and credit-balance lookups). The \textasciitilde{}13k delta against PaymasterV4 isolates the additional cost of SuperPaymaster's SBT eligibility and credit-state logic within the AOA family. This trace-level attribution is drawn from representative transactions on Optimism Mainnet (see Appendix D for transaction hashes).

\subsubsection{Contract Dependency Graph}

\begin{figure}[htbp]
  \centering
  \makebox[\textwidth][c]{\includegraphics[width=1.1\textwidth]{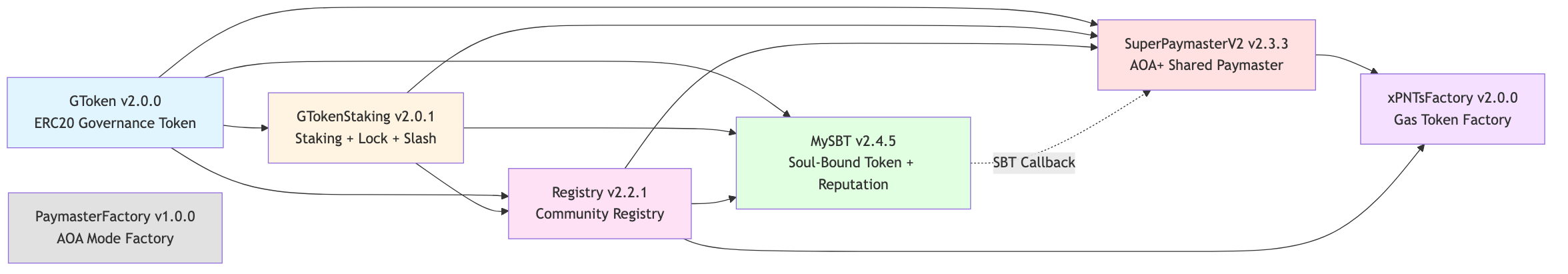}}
  \caption{Contract dependency graph. Arrows indicate read/callback dependencies between on-chain contracts.}
  \label{fig:contract-deps}
\end{figure}

\begin{sloppypar}
The core contracts form a layered logical dependency structure (Figure~\ref{fig:contract-deps}). At the base, the GToken (ERC-20 governance token) feeds into GTokenStaking (staking, lock, and slash logic) and the Registry (community configuration hub). The MySBT contract (Soulbound Token with reputation) depends on GToken, GTokenStaking, and Registry. The SuperPaymaster contract orchestrates gas sponsorship and is logically anchored to all four lower-layer contracts; the xPNTsFactory (gas token factory) is jointly controlled by SuperPaymaster and Registry. To respect the ERC-4337 / ERC-7562 \cite{ERC7562} validation-phase storage rules at runtime, the deployed SuperPaymaster (v3.2.2 on Optimism Mainnet) does not perform live cross-contract SLOADs into these lower-layer contracts during \texttt{validatePaymasterUserOp}. Instead, SBT-eligibility status is mirrored into a \texttt{sbtHolders} mapping inside the paymaster's own storage (written by the Registry via the authorized \texttt{updateSBTStatus} entry point), and the ETH/aPNTs price is mirrored into a \texttt{cachedPrice} struct through a two-tier update path: (1) \textbf{primary}---an on-chain Chainlink oracle feed, the default authoritative source when available; (2) \textbf{fallback}---a DVT-run keeper (three or more independent nodes, permissionless participation, multi-source aggregation from Chainlink, Binance, OKX, and Uniswap V3, with aggregated BLS signatures preventing single-node manipulation) invoked via \texttt{updatePrice}/\texttt{updatePriceDVT} when the Chainlink feed is stale or unavailable. The validation-phase entry point therefore reads only paymaster-self storage. This layered design ensures that each contract has a minimal, well-defined trust surface, while the mirror-and-stake pattern keeps the hot path bundler-compatible. §4.3 provides the per-rule mapping.
\end{sloppypar}

\subsubsection{Core Object Relationships}

\begin{figure}[htbp]
  \centering
  \includegraphics[width=0.68\textwidth]{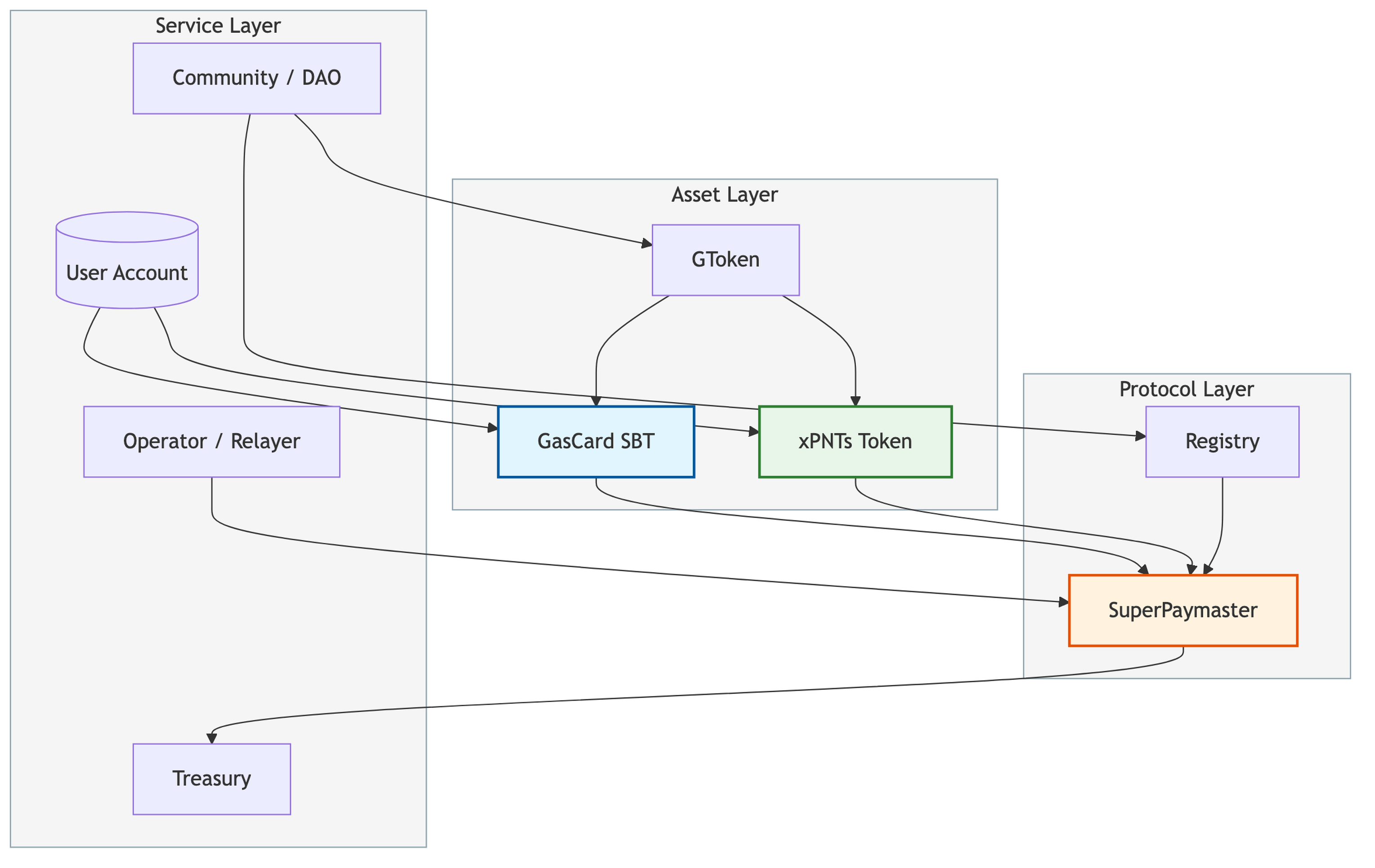}
  \caption{Core object relationships across the Asset, Protocol, and Service layers.}
  \label{fig:core-objects}
\end{figure}

Figure~\ref{fig:core-objects} illustrates the core object relationships across the three architectural layers.

\subsubsection{Community Modes}

\begin{figure}[htbp]
  \centering
  \includegraphics[width=0.85\textwidth]{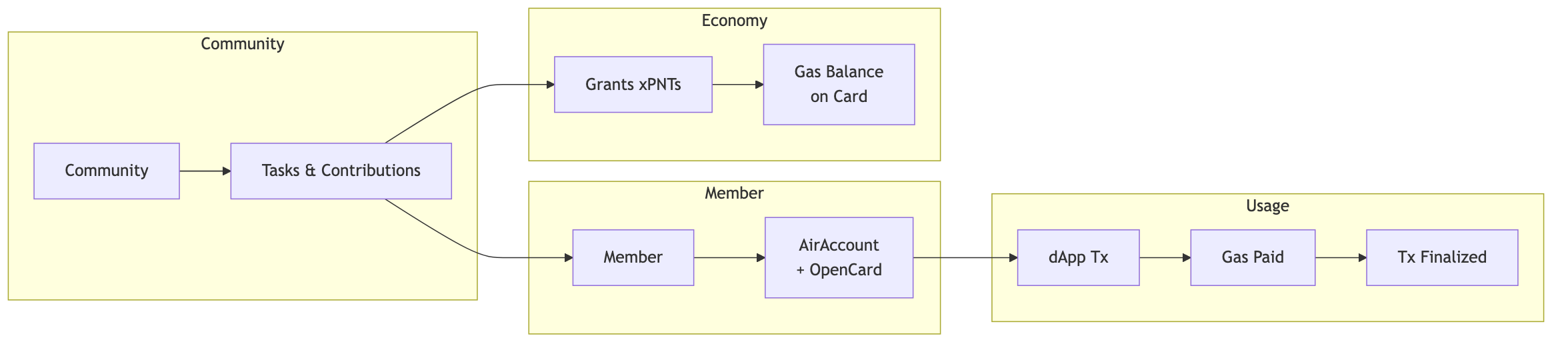}
  \caption{Open Community Mode: community permissionlessly deploys xPNTs + paymaster operator; users separately self-mint the Gas Card SBT from the protocol contract.}
  \label{fig:community-mode}
\end{figure}

SuperPaymaster supports two operational modes. In \textbf{Standard Mode}, the SDK resolves the SuperPaymaster contract and reads operator configuration and supported-token lists from the Registry. In \textbf{Open Community Mode}, any community can permissionlessly deploy xPNTs (community gas tokens) via the xPNTsFactory and register as a paymaster operator in the Registry---without deploying or issuing the Gas Card SBT, which is a protocol-level asset self-minted independently by each user. Figure~\ref{fig:community-mode} shows this flow.

\subsubsection{Economic Circulation}

\begin{figure}[htbp]
  \centering
  \includegraphics[width=0.70\textwidth]{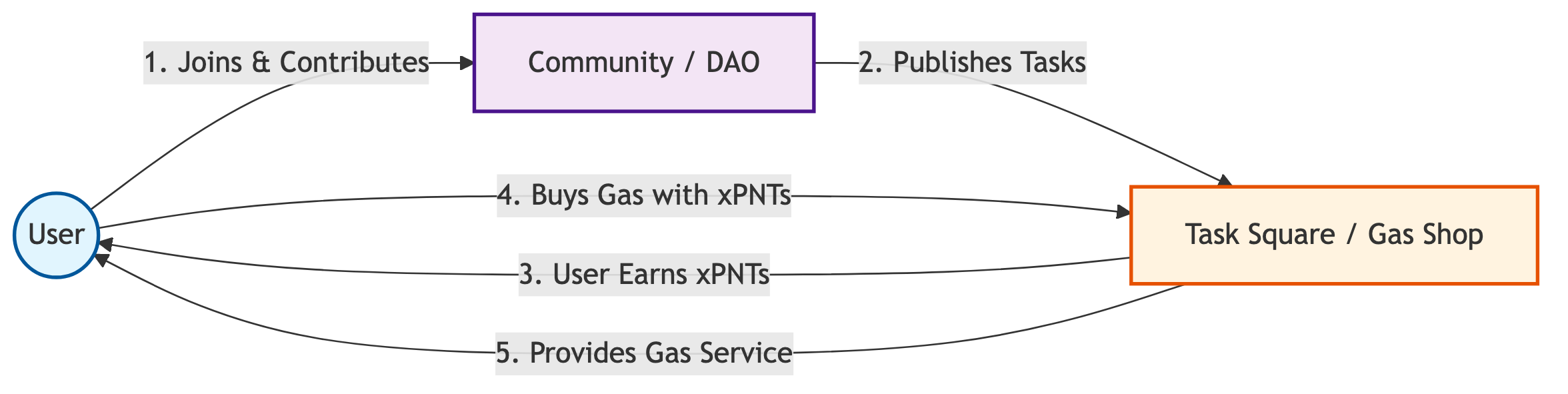}
  \caption{Intended economic circulation: community gas sponsorship enables engagement; whether this creates net mutual value requires empirical field study.}
  \label{fig:economic-circulation}
\end{figure}

Gas sponsorship can create mutual value between communities and their members: communities allocate xPNTs to cover gas costs and may, in return, receive task completion, governance participation, or increased engagement from sponsored users. Whether the value of this engagement exceeds the sponsorship cost depends on community-specific factors and has not been empirically measured; field studies are identified as future work. Figure~\ref{fig:economic-circulation} illustrates this intended circulation.

\subsection{Zero-Approve Gas Token Architecture}

A significant efficiency advantage stems from the xPNTs token's built-in Auto-Approved Spender mechanism. In contrast to commercial ERC-20 paymasters (e.g., Pimlico) that require on-chain \texttt{approve()} plus DEX swap/transfer in \texttt{postOp}, xPNTs eliminates both:

\begin{sloppypar}
\begin{enumerate}\tightlist
\item \textbf{Zero \texttt{approve()} transactions.} \texttt{allowance()} returns \texttt{type(uint256).max} for trusted contracts (SuperPaymaster, Factory, MySBT), removing the separate approval transaction (\textasciitilde{}46,000 gas saved per first interaction).
\item \textbf{Zero on-chain swap.} Commercial ERC-20 paymasters must swap user tokens to ETH via on-chain DEX during \texttt{postOp} (+\textasciitilde{}150,000 gas). SuperPaymaster instead burns the user's xPNTs and transfers the equivalent aPNTs to the protocol treasury in a single atomic \texttt{postOp} step, avoiding any external swap.
\item \textbf{Firewall security.} Despite unlimited allowance, a \texttt{transferFrom()} firewall restricts auto-approved spenders to transfer funds \emph{only} to themselves or to the registered SuperPaymaster address, preventing unauthorized extraction. A per-transaction cap (\texttt{MAX\_SINGLE\_TX\_LIMIT = 5,000 ether} in token units) acts as a circuit-breaker against implementation bugs.
\item \textbf{UserOpHash authentication.} The paymaster's \texttt{validatePaymasterUserOp} receives and verifies the \texttt{userOpHash} of each sponsored operation, binding sponsorship to the exact UserOp and preventing forged or replayed withdrawal attempts.
\end{enumerate}
\end{sloppypar}

The auto-approved spender list is not user-modifiable; adding or removing entries requires an authorized governance action (timelock/multisig). The per-transaction cap is a safety circuit-breaker, not an economic limit; community operators are expected to configure per-card spending caps at the SuperPaymaster policy layer to enforce economically meaningful bounds. This architecture directly explains the gas pattern observed in §5.3: SuperPaymaster's L2 \texttt{txGasUsed} stays consistently near 167,830 gas, while Pimlico's ERC-20 path varies between 226k and 638k gas because of its on-chain token swap.

Table 2 contrasts the two settlement architectures to highlight the structural cost differences that drive the evaluation results in §5.

\begin{table}[htbp]\centering
\resizebox{\textwidth}{!}{%
\begin{tabular}{|l|l|l|}
\hline
\textbf{Dimension} & \textbf{ERC-20 Paymaster (Pimlico-style)} & \textbf{SuperPaymaster (Zero-Approve)} \\
\hline
Token settlement & On-chain DEX swap (Oracle $\rightarrow$ Approve $\rightarrow$ Swap) & Internal SSTORE (account ledger: burn xPNTs + transfer aPNTs to treasury) \\
Gas overhead & \textasciitilde{}100k+ gas for swap sequence & O(1) storage write (\textasciitilde{}5k gas) \\
Price dependency & Real-time DEX liquidity & Operator-set rate (auditable on-chain) \\
User approval & ERC-20 \texttt{approve()} required & No approval needed (Zero-Approve) \\
Censorship surface & DEX + oracle + router contracts & Internal accounting only \\
\hline
\end{tabular}
}
\end{table}

\noindent\textit{Table 2: Architectural comparison of ERC-20 Paymaster (DEX-routed) vs SuperPaymaster (Zero-Approve) settlement paths.}

\subsection{Design Alternatives Considered and Rejected}

Before arriving at the SBT-based Gas Card architecture, we considered and rejected three alternatives, consistent with Hevner's Guideline 6 (Design as a Search Process) \cite{hevner2004design}:

\textbf{(a) ERC-20 Voucher Tokens.} A transferable ERC-20 voucher representing prepaid gas credits was rejected because transferable vouchers create secondary markets that incentivize Sybil farming (mint vouchers via fake accounts, sell on DEX). Non-transferability (SBT) eliminates this attack vector while preserving the asset-ownership mental model.

\textbf{(b) Atomic On-Chain Swap at Validation Time.} Performing a real-time token-to-ETH swap within \texttt{validatePaymasterUserOp} was rejected because DEX dependency adds \textasciitilde{}100k gas overhead per operation and introduces oracle manipulation and liquidity risks. This is the path taken by Pimlico-style ERC-20 paymasters, and our evaluation (§5.3) empirically demonstrates the gas cost of this approach.

\textbf{(c) Synchronous Off-Chain Settlement with On-Chain Fallback.} A design where the paymaster first attempts off-chain settlement (via a trusted settlement server) and falls back to on-chain settlement on failure was rejected because it reintroduces the off-chain signer gate that AOA explicitly aims to remove. Any design where the "happy path" depends on an off-chain server preserves the censorship surface that motivates this research.

These rejections shaped the final architecture: an on-chain SBT for identity binding, an internal account ledger for atomic settlement (burn xPNTs + aPNTs transfer to protocol treasury), and UserOpHash-bound sponsorship for replay safety.

\subsection{User and Developer Journeys}

\begin{figure}[htbp]
  \centering
  \includegraphics[width=0.85\textwidth]{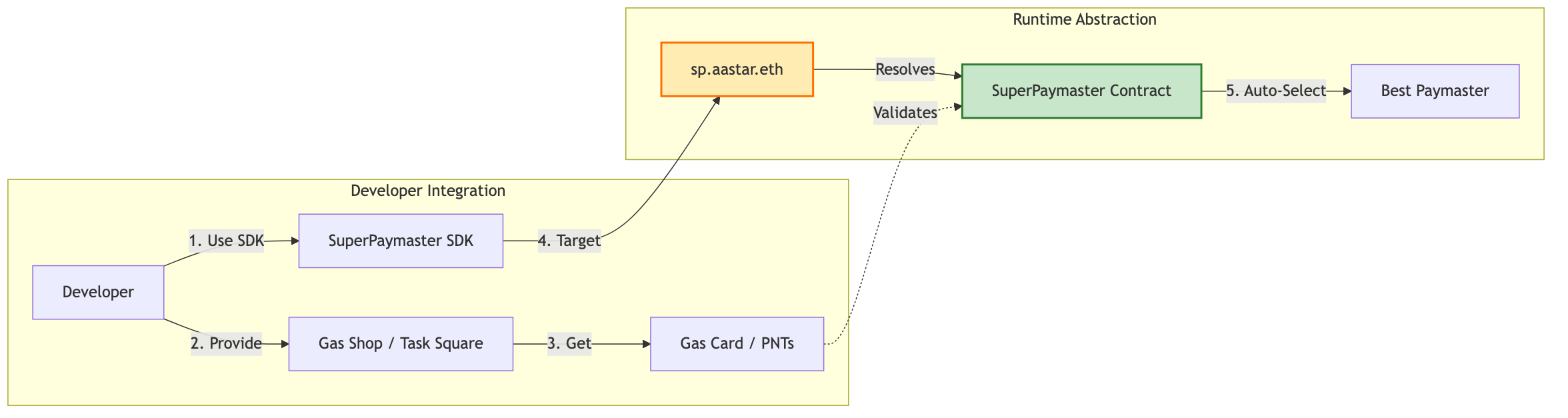}
  \caption{Developer journey: SDK integration simplifies paymaster selection.}
  \label{fig:developer-journey}
\end{figure}

For end users, the SuperPaymaster journey collapses traditional friction into a one-time setup plus single-tap execution: the user claims a Gas Card once, then interacts with any dApp that supports a standard ERC-4337 paymaster; the wallet automatically pays gas from the Card and the transaction confirms. There are no swaps, no bridges, and no \emph{insufficient ETH} errors---complexity is absorbed by the artifact. For developers, integration requires only the SDK, which resolves the SuperPaymaster contract and auto-selects a paymaster; the complexity of paymaster selection is handled on-chain. Figure~\ref{fig:developer-journey} illustrates the developer integration path.

\subsection{Threat Model}

The SuperPaymaster design addresses six threat vectors within its scope.

\textbf{(i) Sybil Attacks on Gas Card Issuance.} Sybil attacks are mitigated by SBT non-transferability: because Gas Cards cannot be transferred or sold, an attacker cannot accumulate sponsorship capacity across fake accounts without creating and funding each account independently. Each SBT registration requires burning a small amount of GToken (approximately \textyen{}0.3--3 per registration under typical GToken prices), establishing a protocol-level cost floor for batch account creation. Applications built on the protocol can raise this floor further by imposing additional eligibility criteria---such as staking requirements, referral gates, or task completion---without modifying the protocol layer.

\textbf{(ii) Gas Card Theft or Compromise.} The SBT is non-transferable by design, so payment eligibility is immutably bound to the owner's Smart Account. Key recovery is provided by the account's programmable spending controls and social-recovery mechanisms (outside this paper's scope). An attacker who compromises the account key gains access to the Card, but this is equivalent to general account compromise and is mitigated at the wallet layer.

\begin{sloppypar}
\textbf{(iii) Paymaster Fund Drain.} Malicious or buggy operations that attempt to drain the paymaster's ETH deposit are constrained by \texttt{MAX\_SINGLE\_TX\_LIMIT} and operator-level circuit breakers that halt sponsorship on anomalous patterns. The per-transaction cap bounds worst-case exposure per operation.
\end{sloppypar}

\textbf{(iv) Centralized Sequencer Censorship.} An L2 sequencer refusing inclusion is considered out of scope; the design inherits the underlying L1/L2 security assumptions and treats sequencer-level inclusion as a separate concern (Section 6.1.3). Where the underlying rollup exposes a force-inclusion mechanism---e.g., OP Stack's L1 deposit transactions, which allow an end user to force a UserOp into the L2 chain via the L1 \texttt{OptimismPortal} contract---a censored user retains a non-discretionary fallback path. This path requires no new paymaster-side signature, preserving the AOA censorship-resilience property at the rollup layer.

\textbf{(v) Registry Governance and SBT Issuance.} The Registry manages two distinct governance surfaces that must not be conflated. First, the \emph{operator} whitelist governs per-community configuration rights (rate limits, credit parameters, operator keys)---this surface is multi-sig- and timelock-governed. Second, Gas Card SBT issuance is \textbf{permissionless} at the protocol layer: any user may mint a single SBT by burning approximately \textyen{}0.3--3 in GToken\footnote{The \textyen{}0.3--3 range is an indicative estimate based on GToken's initial community governance pricing; the exact burn amount per registration is a community-governed parameter adjustable via multi-sig. Applications built on the framework may require additional eligibility criteria beyond the protocol-level burn, raising the effective Sybil cost floor.}; the MySBT contract enforces eligibility entirely on-chain, not through a server gatekeeping API. These two surfaces are independent---operator whitelist membership does not gate SBT minting. In POA, the off-chain signer is a \emph{hard gate}: no transaction is valid without its per-request approval. In AOA, the corresponding governance risk is Registry capture; however, because SBT issuance is permissionless and users register communities on a single SBT rather than depending on any one operator, no single registry operator controls all users' eligibility---if one community is adversarial, users can associate their SBT with any other participating community. The off-chain paymaster signer gate is therefore architecturally eliminated. At current deployment maturity, issuer plurality is aspirational rather than achieved (see Limitation~(iii) in §6.4). The validity-gate elimination claim, however, rests on a technical fact independent of operator breadth: SuperPaymaster replaces the off-chain server signature with on-chain SLOAD verification, which is structurally enforced regardless of how many community operators exist. The protocol also provides a fully self-deployable alternative: the PaymasterV4 deployment mode requires no external community or AAStar infrastructure at all---any technically capable team can permissionlessly deploy MySBT + PaymasterV4 and achieve complete independence from shared operators. For communities without technical capacity who choose SuperPaymaster's shared infrastructure, AAStar's operational role is fork-replaceable: all contracts are open-source (Apache 2.0), so any party can independently replicate the full service stack, creating a fork-equilibrium that disciplines operator behavior. The ``issuer plurality'' limitation refers to the breadth of active independent operators, not to the technical availability of the self-deployable path.

\textbf{(vi) xPNTs Burn Safety.} Because \texttt{postOp} atomically burns user xPNTs and transfers aPNTs to the protocol treasury, a malicious actor who somehow triggers \texttt{postOp} outside a valid UserOp context could attempt unauthorized burns. Three composed mitigations ensure that no xPNTs can be burned without a validly signed and verified UserOperation:

\begin{itemize}\tightlist
\item \textbf{Caller restriction.} The \texttt{postOp} function is callable only by the ERC-4337 EntryPoint contract (enforced by the inherited \texttt{BasePaymaster} access control), so external addresses cannot invoke it directly.
\item \textbf{UserOp binding.} Each sponsorship is bound to a specific \texttt{userOpHash} that is verified during \texttt{validatePaymasterUserOp}, ensuring that \texttt{postOp} settles only the operation it was authorized to settle.
\item \textbf{Per-tx ceiling.} \texttt{MAX\_SINGLE\_TX\_LIMIT} caps the xPNTs that any single sponsored UserOp can burn, bounding the loss of any individual misbehaving operation.
\end{itemize}

\section{Evaluation}

This chapter reports empirical and analytical results from our DSR evaluation activity.

\subsection{Paymaster-Signer Decentralization (validates P1)}

P1 targets a specific centralization layer: the off-chain paymaster signing server that POA services require for every sponsored transaction. BundleBear data \cite{BundleBear2024} shows a single provider processes 71\% of all sponsored UserOperations; if that provider's signing server refuses service or goes offline, those transactions cannot proceed---the signature is a hard validity gate at the ERC-4337 \texttt{EntryPoint} level.

\begin{sloppypar}
\textbf{Architectural evidence.} SuperPaymaster's \texttt{validatePaymasterUserOp} performs zero off-chain calls. All eligibility reads are on-chain SLOAD operations confined to the paymaster's own storage, satisfying the validation-phase self-storage rule formalized in ERC-7562 \cite{ERC7562} (STO-021 in the published rule index). Specifically, the deployed contract reads (i) operator configuration in \texttt{operators[operator]}; (ii) per-(operator, sender) state in \texttt{userOpState}; (iii) the cached price oracle in \texttt{cachedPrice}; and (iv) the SBT eligibility set in \texttt{sbtHolders}---all four reside in the paymaster's own storage. Cross-contract dependencies on Registry, MySBT, GTokenStaking, and Chainlink are kept up to date by authorized writes outside the hot path: \texttt{Registry.updateSBTStatus} mirrors community SBT changes, while \texttt{updatePrice}/\texttt{updatePriceDVT} mirror Chainlink (or DVT-aggregated BLS) feeds. The paymaster itself is staked at the EntryPoint (via \texttt{addStake} in \texttt{BasePaymaster}), which is the precondition under which the self-storage rule applies. This mirror-and-stake pattern is what makes the layered logical architecture of §4.3 bundler-compatible at runtime. The trust-dependency shift is summarized in Table 3:
\end{sloppypar}

\begin{table}[htbp]\centering
\resizebox{\textwidth}{!}{%
\begin{tabular}{|l|l|l|}
\hline
\textbf{Layer} & \textbf{POA (Alchemy-style)} & \textbf{AOA (SuperPaymaster)} \\
\hline
Sponsorship validity & Off-chain API signature (\textbf{hard gate}) & On-chain SBT + policy state \\
Bundler/Relayer & Inclusion only (switchable) & Inclusion only (switchable) \\
Sequencer & L2 ordering (out of scope) & L2 ordering (out of scope) \\
\hline
\end{tabular}
}
\end{table}

\noindent\textit{Table 3: Trust-dependency comparison. AOA moves the sponsorship validity gate on-chain; bundler and sequencer layers are identical in both models.}

Because sponsorship validity is determined entirely by contract state, any standard ERC-4337 bundler can process a SuperPaymaster-sponsored UserOp---the user is not bound to a specific service provider.

\begin{sloppypar}
\textbf{Code structural contrast.} The architectural elimination of the off-chain signing gate is most directly visible at the contract level. In a representative POA paymaster (e.g., Alchemy Gas Manager), \texttt{validate\-Paymaster\-User\-Op} follows the standard ERC-4337 verifying-paymaster pattern\footnote{Canonical reference: \texttt{eth-infinitism/\allowbreak{}account-abstraction}, \texttt{contracts/\allowbreak{}samples/\allowbreak{}VerifyingPaymaster.sol}. The pattern calls \texttt{ECDSA.recover()} on a hash of the \texttt{UserOperation} signed by a server-controlled key; a representative on-chain instance is in Appendix~D (B1\_Alchemy); the sampled Alchemy Gas Manager paymaster contract on Optimism Mainnet is \texttt{0x4Fd9098af9ddcB41DA48A1d78F91F1398965addc}; readers can independently verify that this contract implements the ERC-4337 VerifyingPaymaster (ECDSA-based off-chain signing) pattern via Optimism Etherscan (\texttt{optimistic.etherscan.io}).}: it calls \texttt{ECDSA.recover()} on a signature packed into \texttt{paymaster\-AndData}---a signature that must be freshly generated by an off-chain server for each request. If that server refuses service or blacklists the sender, \texttt{validate\-Paymaster\-User\-Op} returns an error regardless of which bundler forwards the UserOp; the gate is at the \emph{validity} layer, not the inclusion layer. Switching bundlers cannot bypass it because the missing signature is a precondition enforced by the \texttt{EntryPoint} itself. SuperPaymaster's \texttt{validate\-Paymaster\-User\-Op} performs no \texttt{ecrecover} on any off-chain key. Eligibility is determined by four on-chain SLOAD operations: \texttt{operators[operator]}, \texttt{userOpState\allowbreak{}[operator][sender]}, \texttt{cachedPrice}, and \texttt{sbtHolders[sender]}---all confined to the paymaster's own storage per ERC-7562 STO-021 \cite{ERC7562}. No off-chain server participates in, or can veto, this read. \emph{Empirical corroboration}: the N\,=\,50 Optimism Mainnet transactions collected per the protocol in §3.1.1 (Workflow C, T2.1) were submitted and confirmed on-chain without any discretionary paymaster-server approval step, providing direct operational evidence of end-to-end function independent of a signing server. Representative transaction hashes are in Appendix~D; the full per-transaction dataset is archived in the pinned \texttt{aastar-sdk} repository (commit \texttt{03d0ca9}; see Data Availability).
\end{sloppypar}

\textbf{Scope.} This evidence validates that the off-chain paymaster signer is architecturally eliminated as a validity gate. It does not claim unconditional censorship resistance across all layers: bundler selection, RPC access, and sequencer ordering remain separate inclusion surfaces (§6.1.3). Adversarial scenarios with sustained, coordinated attacks remain future work. \emph{Evidence supports P1.}

\subsection{Analytical Usability via GOMS (validates P2)}

We employ the GOMS model (Goals, Operators, Methods, Selection rules) \cite{Card1983} as an analytical evaluation of interaction complexity. Following Card et al.'s operator taxonomy, \textbf{M} (Mental operator) denotes a cognitive step --- perceiving, deciding, or recalling (\textasciitilde{}1.35 s); \textbf{P} (Physical/Pointing operator) denotes a motor action --- clicking, navigating, or executing (\textasciitilde{}1.10 s); and \textbf{W} (Wait operator) denotes system-imposed latency such as confirmation delays (variable duration). Operator counts serve as an interaction-complexity proxy, not precise timing predictions.

\textbf{Model A (Traditional EOA, failure-recovery path).} A user encountering "Insufficient ETH" must: (M) perceive error; (M) decide strategy (bridge/swap/buy); (M) select external service; (P) navigate to it; (M) calculate amount needed under gas-price volatility; (P) execute swap/bridge; (W) wait for confirmation; (P) return to the dApp; (P) retry. \textbf{Total: 4M + 4P + 1W = 9 operators. Cognitive load: High.}

\textbf{Model C (POA paymaster, steady-state).} A user transacting via a vendor-operated paymaster must: (M) perceive transaction prompt; (M) decide which paymaster service to use; (P) initiate and configure the paymaster API call; (P) sign and submit via the paymaster. \textbf{Total: 2M + 2P = 4 operators. Cognitive load: Moderate} (requires trust judgment toward a third-party off-chain signer).

\textbf{Model B (AOA Gas Card) --- two phases.} The AOA model distinguishes a one-time setup phase from the recurring per-transaction phase.

\emph{Phase 0 --- One-time initialization (performed once; cost amortized across all subsequent transactions):} (M) decide to join a community and evaluate available communities; (P) register as a community member by claiming an SBT on-chain; (W) wait for SBT mint confirmation; (P) load the Gas Card by receiving community-distributed xPNTs or performing a self top-up. \textbf{Init subtotal: 1M + 2P + 1W = 4 operators.}

\emph{Phase 1 --- Steady-state per transaction:} (M) perceive transaction prompt; (P) click "Confirm" --- Gas Card auto-deducts xPNTs. \textbf{Per-transaction subtotal: 1M + 1P = 2 operators. Cognitive load: Minimal.}

\emph{Phase 2 --- Occasional top-up (when balance is low):} (M) perceive low-balance notification; (P) recharge the Gas Card (purchase or claim additional xPNTs). \textbf{Top-up subtotal: 1M + 1P = 2 operators} --- equivalent to one normal transaction, and significantly lighter than EOA's recurring 9-operator failure-recovery path.

Table 4 enumerates all operators step-by-step across the three models.

\begin{table}[htbp]\centering
\resizebox{\textwidth}{!}{%
\begin{tabular}{|l|l|l|l|}
\hline
\textbf{Step} & \textbf{EOA-direct (Model A)} & \textbf{POA paymaster (Model C)} & \textbf{AOA Gas Card (Model B)} \\
\hline
\textbf{[ONE-TIME SETUP]} & --- & --- &  \\
I-1 & --- & --- & M --- decide to join a community \\
I-2 & --- & --- & P --- register as member (claim SBT on-chain) \\
I-3 & --- & --- & W --- wait for SBT mint confirmation \\
I-4 & --- & --- & P --- load Gas Card (receive/purchase xPNTs) \\
\textbf{Init subtotal} & --- & --- & \textbf{1M + 2P + 1W = 4 (one-time)} \\
\textbf{[PER TRANSACTION]} &  &  &  \\
1 & M --- perceive "Insufficient ETH" error & M --- perceive transaction prompt & M --- perceive transaction prompt \\
2 & M --- decide strategy (bridge/swap/buy) & M --- decide which paymaster to use & P --- click Confirm (auto-deducts xPNTs) \\
3 & M --- select external service & P --- configure paymaster API call & --- \\
4 & P --- navigate to external service & P --- sign \& submit via paymaster & --- \\
5 & M --- calculate amount (gas volatility) & --- & --- \\
6 & P --- execute swap / bridge & --- & --- \\
7 & W --- wait for confirmation & --- & --- \\
8 & P --- return to dApp & --- & --- \\
9 & P --- retry original transaction & --- & --- \\
\textbf{Per-tx total} & \textbf{4M + 4P + 1W = 9} & \textbf{2M + 2P = 4} & \textbf{1M + 1P = 2} \\
\textbf{[OCCASIONAL TOP-UP]} & (same 9-operator path, recurring) & (\textasciitilde{}2--3 operators, config refresh) & M --- perceive low-balance notice \\
 &  &  & P --- recharge xPNTs \\
\textbf{Top-up subtotal} & \textbf{9 (recurring failure path)} & \textbf{\textasciitilde{}2--3} & \textbf{1M + 1P = 2} \\
\hline
\end{tabular}
}
\end{table}

\noindent\textit{Table 4: GOMS operator-level decomposition across the three workflow models---Model A (EOA-direct, failure-recovery path), Model B (AOA Gas Card), and Model C (POA paymaster)---covering one-time setup, per-transaction steady state, and occasional top-up.}

\begin{figure}[htbp]
  \centering
  \includegraphics[width=0.85\textwidth]{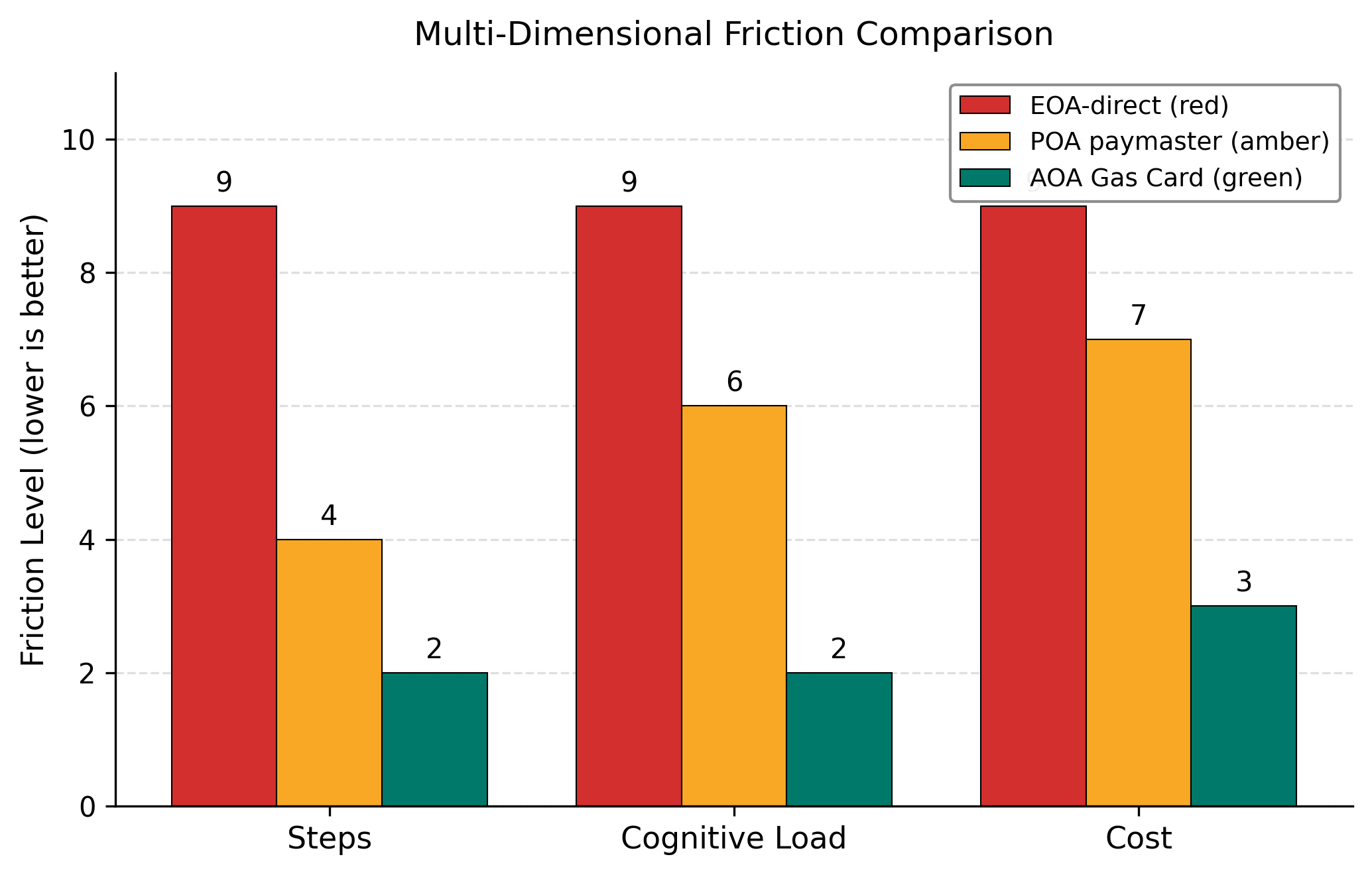}
  \caption{Multi-dimensional friction comparison across three workflows (Steps, Cognitive Load, Cost; lower is better), per-transaction steady-state path. All scores are ordinal and not ratio-scale. EOA-direct (failure-recovery bound, red): Steps=9 (GOMS count), Cognitive Load=9, Cost=9. POA paymaster (amber): Steps=4 (GOMS count), Cognitive Load=6, Cost=7. AOA Gas Card (green): Steps=2 (GOMS count), Cognitive Load=2, Cost=3. Steps maps directly to GOMS per-transaction operator count; Cognitive Load and Cost are ordinal estimates illustrating directional reduction. EOA 9-operator path is a recurring worst-case bound, not a steady-state comparison. AOA one-time setup cost acknowledged in Table~4 and amortized across subsequent transactions.}
  \label{fig:friction-comparison}
\end{figure}

\noindent\textit{Note: M = Mental operator (perceive, decide, recall; \textasciitilde{}1.35 s per Card et al. \cite{Card1983}); P = Physical/Pointing operator (click, navigate, execute; \textasciitilde{}1.10 s); W = Wait operator (system-imposed latency, variable duration). Operator counts are simplified step counts, not full timing predictions. The multi-dimensional friction comparison (Figure~\ref{fig:friction-comparison}) compares the per-transaction steady-state path only; the AOA one-time setup cost is explicitly acknowledged above and amortized across all subsequent transactions.}

In the per-transaction \textbf{steady-state} path, AOA (Model B) requires \textbf{2 operators} (1M + 1P) versus POA (Model C) at \textbf{4 operators} (2M + 2P)---a 50\% reduction. POA steady-state requires the user to select a paymaster service and form a trust judgment toward an off-chain signer; AOA eliminates both steps via persistent on-chain eligibility. The EOA failure-recovery path (Model A, 9 operators: 4M + 4P + 1W) is a \emph{recurring} upper bound: every gas-insufficient transaction triggers the full recovery sequence, making the 9-to-2 reduction (78\%) the maximum modeled improvement across gas-failure scenarios. AOA's 4-operator one-time setup cost is amortized across all subsequent 2-operator transactions. Eliminated per-transaction operators (service selection, amount calculation, swap execution, wait, return) are gas-management overhead extraneous to the user's core intent---precisely the extraneous cognitive load that Sweller's CLT prescribes eliminating \cite{Sweller1988}.

\textbf{Scope and limitations.} Model A captures the failure-recovery path (users acquiring gas tokens before retry); Model B captures steady-state usage after one-time Gas Card setup. Model C captures steady-state usage with an always-available vendor paymaster (specifically, a user-visible paymaster selection scenario representative of multi-chain or cross-dApp contexts where a single provider does not cover all target chains). If the user already holds the required payment assets in both Model A and Model B, the EOA workflow also reduces to 1M + 1P; the GOMS comparison therefore illustrates the advantage of eliminating \emph{recurring} gas-management operators, not a universal superiority claim. AOA's steady-state advantage is structural rather than cosmetic. Even in deeply integrated wallet deployments where paymaster selection is invisible, POA requires that either the user holds ETH or relies on a central operator for per-transaction sponsorship---neither option is portable across heterogeneous chains and dApps, nor free of service-availability dependency. Under AOA, the Gas Card is a persistent on-chain asset the user owns: no recurring ETH balance is required per transaction, and eligibility is enforced by deterministic on-chain state regardless of which chain or dApp the user interacts with. Operator counts are simplified step counts, not full GOMS timing predictions \cite{Card1983}; we do not report operator-level timing estimates. Controlled user studies (SUS, NASA-TLX, task-completion and error rates) remain future work (§6 Future Work). \emph{Evidence supports P2.}

Figure~\ref{fig:friction-comparison} visualizes the multi-dimensional friction comparison across the three workflows.

The multi-dimensional friction comparison figure visualizes GOMS-derived ordinal scores across three dimensions for the per-transaction steady-state path. \textbf{Steps} maps directly to GOMS per-transaction operator count (EOA: 9, POA: 4, AOA: 2). \textbf{Cognitive Load} reflects operator-type composition: workflows dominated by Mental operators (M) impose higher cognitive burden than those requiring only Physical operators (P). \textbf{Cost} is an ordinal estimate of total economic friction: EOA incurs exchange or bridge fees plus gas; POA incurs bundler markup; AOA incurs only an internal xPNTs debit. All three dimensions are ordinal comparisons, not ratio-scale measurements; they illustrate directional reduction across workflows and are not absolute units.

\subsection{Gas Profiling on Optimism Mainnet (validates P3)}

All measurements below are sourced from on-chain \texttt{UserOperationEvent} logs under the controlled conditions defined in §3.1.1; raw data and collection scripts are in Appendix C, transaction hashes for independent verification are in Appendix~D, and representative per-transaction gas-structure decompositions are in Appendix~E.

\begin{figure}[t]
  \centering
  \includegraphics[width=0.75\textwidth]{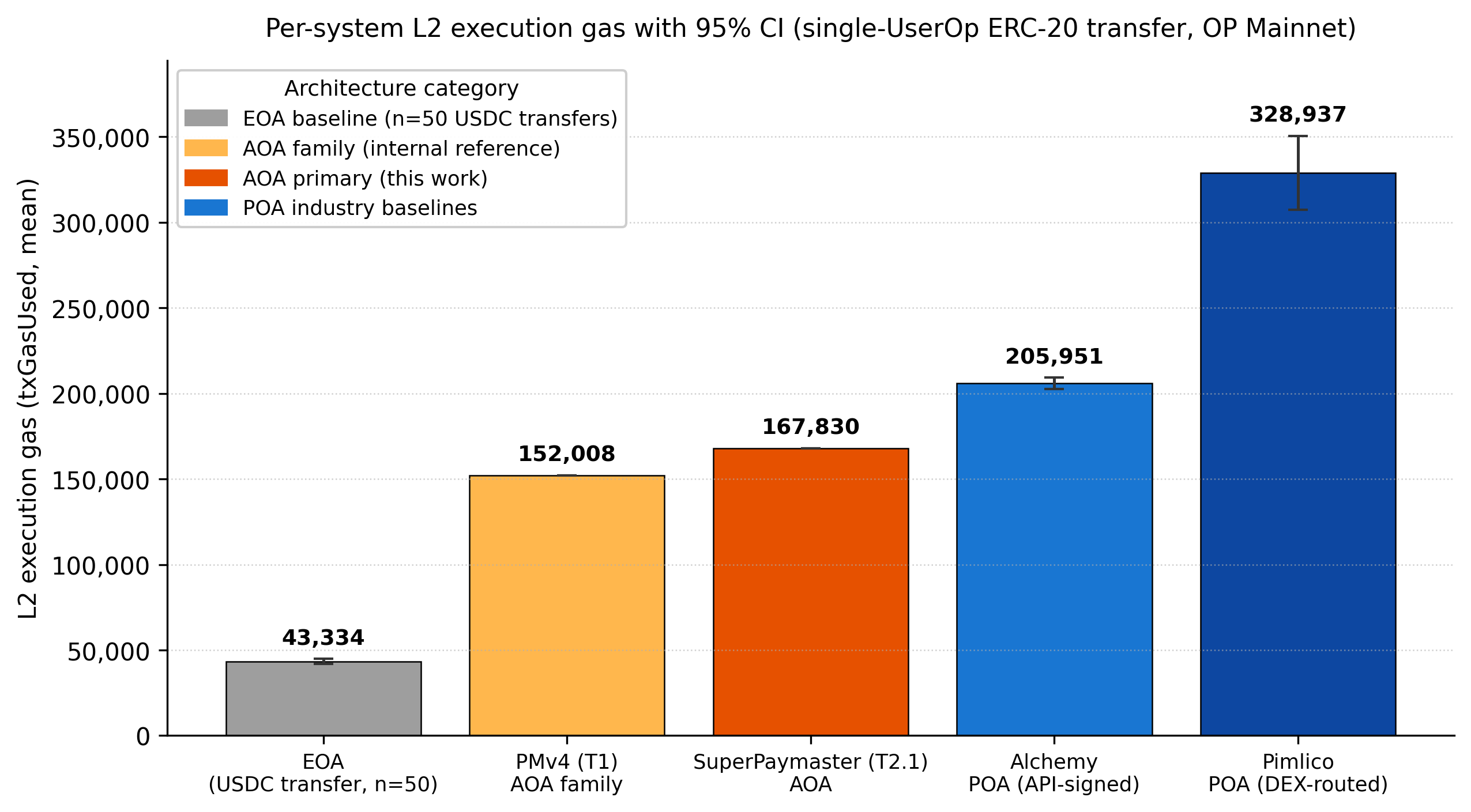}
  \caption{L2 gas used (\texttt{txGasUsed}, mean values) for each operation type. 95\% CIs in Table~5. All operations on OP Mainnet; receipts collected via \texttt{eth\_getTransactionReceipt}.}
  \label{fig:l2-gas-comparison}
\end{figure}

Table 5 summarizes empirical L2 execution gas for each operation type (n = 50 per system, single-UserOp bundles with ERC-20 \texttt{transfer} selector). The metric is \texttt{txGasUsed} (pure on-chain L2 execution gas, stripping bundler PVG so that paymaster-architecture differences are not confounded with bundler pricing); total billed gas (\texttt{actualGasUsed}) and its decomposition appear in Table 6. Figure~\ref{fig:l2-gas-comparison} visualizes the means.

\begin{table}[htbp]\centering
\resizebox{\textwidth}{!}{%
\begin{tabular}{|l|l|r|r|l|}
\hline
\textbf{Operation} & \textbf{Label} & \textbf{n} & \textbf{txGasUsed (mean +/- 95\% CI)} & \textbf{Notes} \\
\hline
EOA ERC-20 Transfer (USDC) & A\_EOA & 50 & 43,334 +/- 1,492 & OP Mainnet direct transfer \\
Gasless Transfer (PaymasterV4) & T1 & 50 & 152,008 +/- 5 & Prepayment mode \\
Gasless Payment (SuperPaymaster Normal) & T2.1 & 50 & 167,830 +/- 7 & Burns community gas token \\
Alchemy (API-Verified) & B1\_Alchemy & 50 & 205,951 +/- 3,222 & EntryPoint v0.6 samples \\
Pimlico (DEX-Routed ERC-20) & B2\_Pimlico & 50 & 328,937 +/- 21,679 & EntryPoint v0.7 samples \\
\hline
\end{tabular}
}
\end{table}

\noindent\textit{Table 5: L2 execution gas (\texttt{txGasUsed}) per operation type with bootstrap 95\% CIs (B=10,000). The metric is pure on-chain execution gas, stripping bundler PVG so that paymaster-architecture differences are not confounded with bundler pricing. Total billed gas (\texttt{actualGasUsed}) and its decomposition appear in Table 6.}

Table 6 decomposes the ERC-4337 billing metric (\texttt{actualGasUsed}) into pure L2 execution gas (\texttt{txGasUsed}) and PVG overhead. Because PVG scales with bundler pricing strategies rather than paymaster architecture, extracting \texttt{txGasUsed} provides a more controlled basis for comparing architectural efficiency.

\begin{table}[t]\centering
\resizebox{\textwidth}{!}{%
\begin{tabular}{|l|l|r|r|r|}
\hline
\textbf{Operation} & \textbf{Label} & \textbf{Billed (\texttt{actualGasUsed})} & \textbf{Execution (\texttt{txGasUsed})} & \textbf{PVG overhead} \\
\hline
EOA ERC-20 & A\_EOA & N/A & 43,334 & N/A \\
PaymasterV4 & T1 & 271,092 & 152,008 & 119,084 \\
SuperPaymaster & T2.1 & 286,818 & 167,830 & 118,988 \\
Alchemy & B1\_Alchemy & 257,299 & 205,951 & 51,348 \\
Pimlico & B2\_Pimlico & 387,129 & 328,937 & 58,192 \\
\hline
\end{tabular}
}
\end{table}

\noindent\textit{Table 6: Gas decomposition. Both SuperPaymaster and Alchemy Gas Manager UserOps were submitted via the Alchemy standard ERC-4337 bundler API, which supports all conformant UserOperations. The PVG disparity (\textasciitilde{}119k for SuperPaymaster vs \textasciitilde{}51k for Alchemy Gas Manager) therefore reflects the Alchemy bundler applying different PVG estimation strategies to different paymaster--account stack combinations, not a difference in bundler software. Differences in \texttt{actualGasUsed} therefore reflect bundler-level PVG pricing as well as paymaster architecture; \texttt{txGasUsed} (pure L2 execution gas) is the more controlled comparison metric.}

\textbf{Primary comparison: AOA vs industry POA baselines.} The central evaluation question is whether AOA can achieve competitive gas efficiency while eliminating the off-chain paymaster signer. We compare SuperPaymaster (n=50) against two representative industry POA paymasters---Alchemy Gas Manager (n=50, API-signature model) and Pimlico ERC-20 paymaster (n=50, DEX-routed model)---all on Optimism Mainnet, filtered for single-UserOp bundles with ERC-20 \texttt{transfer} selector. The industry samples are treated as vendor-as-deployed baselines rather than account-type-controlled experiments. In particular, Alchemy Gas Manager traffic is tightly coupled with Alchemy's account stack, and our chain-level audit found the sampled Alchemy senders to be Alchemy modular/LightAccount-style accounts rather than SimpleAccount instances. We therefore interpret account-type heterogeneity as a residual confound and report both \texttt{actualGasUsed} (ERC-4337 billing) and \texttt{txGasUsed} (pure L2 execution, stripping bundler PVG noise) to separate ERC-4337 billing effects from L2 execution costs. Three structural cost drivers emerge:

\begin{sloppypar}
\begin{enumerate}\tightlist
\item \textbf{SuperPaymaster vs DEX-routed paymasters (Pimlico, mean \texttt{actualGasUsed} 387k; mean \texttt{txGasUsed} 329k).} SuperPaymaster replaces Pimlico's \emph{Oracle $\rightarrow$ Approve $\rightarrow$ Swap} sequence---which alone costs \textasciitilde{}100k+ gas per transaction---with an O(1) internal SSTORE settlement (\textasciitilde{}5k gas). The DEX-liquidation path is an architectural requirement of Pimlico's paymaster model, not a tunable choice; this constitutes a genuine \emph{structural} advantage. In pure L2 execution, SuperPaymaster (168k) is less than half of Pimlico (329k).
\item \textbf{SuperPaymaster vs API-signature paymasters (Alchemy, mean \texttt{actualGasUsed} 257k; mean \texttt{txGasUsed} 206k).} Trace-level decomposition (§4.3.1) isolates the paymaster-specific cost: Alchemy's \texttt{validatePaymasterUserOp} is a \textasciitilde{}16k-gas ECDSA check backed by off-chain signing infrastructure; SuperPaymaster's validation is a \textasciitilde{}48k-gas on-chain state resolution. The \textasciitilde{}32k-gas delta is the measurable execution cost of removing the off-chain signer gate. Despite this heavier validation, SuperPaymaster's \texttt{txGasUsed} (168k) is \emph{lower} than Alchemy's (206k). Because the Alchemy sample reflects its vendor-as-deployed account stack, this comparison should be read as an industry baseline rather than a strict account-type-controlled causal contrast. AOA's structural avoidance of the DEX liquidation path remains the primary architectural advantage, while the validation cost itself is comparable.
\item \textbf{PVG (bundler pricing) explains the remaining gap in \texttt{actualGasUsed}.} SuperPaymaster's \texttt{actualGasUsed} (287k) exceeds Alchemy's (257k) despite lower L2 execution. The difference traces to pre-verification gas (PVG): both systems were submitted via the Alchemy standard ERC-4337 bundler API, yet the bundler applied \textasciitilde{}119k PVG to SuperPaymaster UserOps versus \textasciitilde{}51k to its own Gas Manager UserOps. This disparity reflects the Alchemy bundler applying different PVG estimation strategies to different paymaster--account stack combinations, not a difference in bundler software. PVG is a bundler pricing parameter, not a paymaster architectural property. Appendix E confirms L1 data fees account for \textasciitilde{}8\% of total ETH cost and PVG 31--38\% of billed gas.
\end{enumerate}
\end{sloppypar}

Table 7 maps this onto execution modules. Representative transaction traces (hashes listed in Appendix D) confirm the decomposition.

\begin{table}[htbp]\centering
\resizebox{\textwidth}{!}{%
\begin{tabular}{|l|r|r|l|}
\hline
\textbf{Execution Module} & \textbf{Alchemy (API-Verified)} & \textbf{SuperPaymaster (Normal)} & \textbf{Note} \\
\hline
Account Validation & \textasciitilde{}12,000 & \textasciitilde{}12,000 & SimpleAccount vs Alchemy modular/LightAccount \\
Paymaster Validation & \textbf{\textasciitilde{}16,000} & \textbf{\textasciitilde{}48,625} & On-chain SBT/credit vs. off-chain ECDSA \\
Operation Execution & \textasciitilde{}45,000 & \textasciitilde{}47,000 & ERC-20 transfer \\
EntryPoint + PVG Overhead & \textasciitilde{}184,000 & \textasciitilde{}179,000 & Bundler packaging \\
Total Billed (\texttt{actualGasUsed}) & \textasciitilde{}257,000 & \textasciitilde{}286,000 &  \\
\hline
\end{tabular}
}
\end{table}

\begin{sloppypar}
\noindent\textit{Table 7: Transaction-level trace decomposition. Account-validation costs are measured from on-chain execution traces (\texttt{eth\_trace\_transaction}); the similar values for both account implementations reflect the dominance of secp256k1 ECDSA signature verification in the validation phase, with implementation-specific overheads below the precision of the rounded decomposition.}
\end{sloppypar}

\begin{sloppypar}
\textbf{Internal cost decomposition (PaymasterV4 as same-family reference).} PaymasterV4 is an earlier AOA implementation from the same project that also removes the off-chain signer; it therefore cannot serve as a POA baseline. However, comparing T2.1 (SuperPaymaster, mean \texttt{actualGasUsed} 286,818) against T1 (PaymasterV4, mean 271,092)---using identical EntryPoint v0.7, SimpleAccount, and bundler---isolates the additional gas cost of SuperPaymaster's SBT eligibility, rate-limit, and credit-state logic: +15,726 gas (+5.80\%). Trace-level decomposition attributes this to \texttt{validatePaymasterUserOp} (SuperPaymaster \textasciitilde{}48,625 vs PaymasterV4 \textasciitilde{}35,549; delta \textasciitilde{}13,076 gas). This internal decomposition clarifies \emph{where} the gas goes within the AOA architecture, but the contribution validation rests on the AOA-vs-POA comparison above.
\end{sloppypar}

\textbf{Statistical analysis.} Cliff's $\delta$ for the primary AOA-vs-POA comparisons (computed as $\delta = P(\text{SP} < X) - P(\text{SP} > X)$ so that positive $\delta$ means SP is lower): SuperPaymaster vs Alchemy Gas Manager $\delta = +1.000$ (large); SuperPaymaster vs Pimlico ERC-20 $\delta = +1.000$ (large). Both reflect complete separation---SuperPaymaster's \texttt{txGasUsed} distribution ($\mu = 167{,}830$, $\sigma \approx 430$, range 1,704~gas) does not overlap with either Alchemy ($\mu = 205{,}951$, $\sigma = 3{,}222$) or Pimlico ($\mu = 328{,}937$, $\sigma = 21{,}679$), confirming that the gas differences are architectural rather than stochastic. The complete separation is confirmed from the observed ranges in the collected n=50 samples: SuperPaymaster max $= 167{,}867$ gas $<$ Alchemy observed minimum $= 181{,}234$ gas, so the reported Cliff's $\delta=+1.000$ values reflect non-overlapping empirical distributions rather than a parametric assumption. Cliff's $\delta$ for the AOA-family internal comparison (T1 vs T2.1) is $-1.000$ (large), reflecting complete separation in the reverse direction---SuperPaymaster costs more than PaymasterV4 within the AOA family, with the 15,726-gas delta attributable to SBT eligibility, rate-limit, and credit-state logic. Distribution diagnostics (T1: $\sigma \approx 0$, deterministic; T2.1: $\sigma \approx 430$, range 1,704~gas; T1 skewness $\gamma_1 = 0.70$; T2.1: $\gamma_1 = 0.02$) support means as reliable estimates. The PVG-and-L1-fee decomposition is provided in Appendix E; L1 data fees account for \textasciitilde{}8\% of total ETH cost under the current post-Dencun blob regime \cite{UltraSoundMoney2024}, and PVG (bundler overhead) is 31--38\% of billed gas.

\textbf{Key finding on efficiency vs. decentralization.} SuperPaymaster demonstrates that on-chain sponsorship validation is economically viable relative to industry POA paymasters. The \textasciitilde{}32k gas validation overhead versus ECDSA-based POA is the bounded, predictable cost of eliminating the centralized signer---a trade-off the architecture explicitly pays. Against DEX-routed POA, SuperPaymaster achieves lower L2 execution cost by replacing the \emph{Oracle $\rightarrow$ Approve $\rightarrow$ Swap} sequence with an O(1) internal SSTORE. Beyond on-chain gas, SuperPaymaster converts gas payment into a programmable asset layer (Gas Card + community gas token), enabling invisible payment, policy-controlled sponsorship, predictable interaction under fee volatility, and a community value loop where gas spend can be funded by community incentives rather than forcing users to acquire ETH. \emph{Evidence supports P3.}

\subsection{Qualitative Analysis: Trust Architecture}

Moving from POA to AOA relocates the trust boundary at the sponsorship layer. In POA the flow is User $\rightarrow$ off-chain server (API) $\rightarrow$ blockchain, requiring trust that the server will not censor specific addresses or refuse to sign \cite{Sun2025}. In AOA the flow is User $\rightarrow$ on-chain asset (Gas Card) $\rightarrow$ blockchain; validation is deterministic and requires no off-chain signature. Relayers and bundlers still influence inclusion, but they cannot retroactively invalidate a UserOp that satisfies on-chain eligibility. This narrows the trust boundary at the sponsorship layer while leaving inclusion-layer surfaces (bundler selection, sequencer ordering) as separate concerns (§6.1.3).

\subsection{Summary}

The evaluation indicates that the asset-oriented paymaster design can (i) remove discretionary paymaster-signer authority from sponsorship validity (P1), supported by code structural analysis and on-chain Mainnet evidence; (ii) reduce modeled interaction complexity in gas-payment workflows (P2), supported analytically via GOMS rather than empirical user-study evidence; and (iii) achieve competitive gas efficiency against industry POA baselines---lower L2 execution gas than both API-signature and DEX-routed paymasters---while paying a bounded \textasciitilde{}32k gas validation overhead as the measurable cost of on-chain eligibility verification (P3), supported by mainnet gas profiling (n = 50 per system).

\section{Discussion}

\subsection{Interpretation of Findings}

Our evaluation provides evidence that the asset-oriented paymaster design addresses three dimensions of the research problem---\emph{comprehensive economic cost}, \emph{modeled cognitive load}, and \emph{centralized paymaster-signer authority}---with distinct but complementary evidence types.

\subsubsection{Reducing Economic Cost (Comprehensive, Not Only Gas Units)}

SuperPaymaster's primary advantage is not minimizing raw on-chain execution gas relative to the EOA physical floor, but removing recurring categories of user friction: preparatory cost (acquiring ETH, navigating fee markets) and dApp-specific setup cost (token approvals, per-application gas tanks) common in session-scoped POA models. The gas profiling shows that this friction removal comes at a bounded architectural cost. Trace-level decomposition isolates a \textasciitilde{}32k gas validation overhead versus ECDSA-based POA paymasters---the measurable price of on-chain eligibility verification. Yet in pure L2 execution (\texttt{txGasUsed}), SuperPaymaster achieves lower gas than both evaluated POA paymasters (168k vs Alchemy 206k vs Pimlico 329k), because AOA eliminates the DEX liquidation path and uses a simpler account structure. The remaining \texttt{actualGasUsed} gap versus Alchemy traces to bundler PVG pricing, not paymaster architecture. This evidence suggests that the AOA paradigm can be economically viable while structurally eliminating the off-chain signer dependency.

\subsubsection{Reducing Cognitive Load via Metaphors and Invisible Execution}

The Gas Card metaphor, supported by the GOMS analytical evaluation (§5.2), addresses RQ2 by transforming an abstract, recurring technical process into a tangible, persistent capability. This shift from \emph{process} to \emph{asset} reduces the need for repeated context-dependent setup and decision-making. Beyond the visible metaphor, SuperPaymaster embodies Norman's \textbf{principle of invisibility} \cite{Norman2013}: the best interaction is no interaction. By moving gas negotiation, relayer selection, and payment verification to the protocol layer (the "underwater" part of the iceberg), the user's decision path collapses to a single confirmation. This invisibility is what minimizes \emph{extraneous cognitive load}, allowing users to focus on their primary intent. GOMS evidence is analytical; large-scale user studies (NASA-TLX, SUS, error rates) remain necessary before claiming subjective satisfaction outcomes and are discussed as future work.

\subsubsection{Narrowing the Trust Boundary at the Sponsorship Layer}

\begin{sloppypar}
Decentralization is multi-surface; SuperPaymaster targets a durable bottleneck---paymaster-server signature authority as a validity gate. Code structural analysis demonstrates that \texttt{validatePaymasterUserOp} performs no \texttt{ecrecover} on any off-chain key: eligibility reads are confined to on-chain storage, so no centralized server can veto a transaction that holds a valid Gas Card and sufficient balance. The N\,=\,50 Mainnet transactions in Appendix~D corroborate this architecture empirically---each was confirmed without any discretionary paymaster-server approval. This does not imply unconditional censorship resistance in the full pipeline: bundler selection, RPC access, sequencer ordering, and client middleware remain potential inclusion bottlenecks. However, bundler/relayer censorship is bypassable in principle (users can switch bundlers or submit directly to the \texttt{EntryPoint} via RPC; median Ethereum transaction confirmation times are in the 5--30\,s range \cite{Pacheco2023}, so the overhead of switching is modest relative to normal inclusion latency), whereas paymaster-signer censorship is not---it is the one gate that cannot be bypassed by simply choosing a different relay endpoint. AOA eliminates that gate architecturally.
\end{sloppypar}

\subsection{Theoretical Implications: The Asset-Oriented Paradigm}

The most significant conceptual contribution is the conceptualization of \textbf{Asset-Oriented Abstraction (AOA)} as a design principle for Web3 UX. POA places a significant cognitive burden on users because sponsorship state must be re-established per process. AOA reifies that ephemeral process into a persistent, tangible \emph{asset}; instead of initiating a process to pay for gas, the user \emph{possesses} a Gas Card that \emph{affords} paying for gas. This aligns with Norman's principle of affordance \cite{Norman2013} and bridges the Web2/FinTech mental model to Web3/Crypto implementation. According to Sweller's Cognitive Load Theory \cite{Sweller1988}, POA overloads working memory with transient state (pending approvals, current gas market, per-dApp token lists); AOA offloads this state onto the asset itself---the user checks "Card balance > 0" rather than reconstructing the process state. This is \emph{cognitive offloading} \cite{Risko2016, Clark1998}, not elimination: the user retains a minimal monitoring task (balance awareness), which is substantially lower than multi-step process management.

\textbf{Cognitive offloading (CLT) \cite{Risko2016, Clark1998}.} AOA achieves cognitive offloading by delegating the gas-state tracking task to the Card itself: the user monitors a balance rather than managing a multi-step gas-acquisition process. This offloading is structurally tied to the persistent, user-owned asset model and does not require any off-chain server participation. \textbf{Sovereignty preservation (independent claim).} Independently, AOA preserves sponsorship sovereignty: the user does not delegate authority over \emph{who may transact} to a centralized server. These two claims are distinct. A POA paymaster could achieve cognitive offloading (auto-approval, invisible UX) while still retaining a centralized signer---it would score well on the offloading dimension but fail on sovereignty. AOA achieves both simultaneously. The on-chain eligibility check incurs a \textasciitilde{}32k gas validation overhead compared to ECDSA-based off-chain signing in POA paymasters (§5.3); this is the measurable cost of preserving sovereignty rather than delegating it. It is also worth distinguishing the paymaster layer from the inclusion layer: even under on-chain sponsorship validity, inclusion still depends on bundler/sequencer behavior under ERC-4337. The Ethereum roadmap is actively absorbing the bundler into the execution layer itself, so over time that surface ceases to be an independent censorship bottleneck. The paymaster role---deciding \emph{who pays} and \emph{under what conditions}---remains a persistent economic role regardless of native AA, which is why decentralizing the sponsorship validity gate is the more durable contribution.

\subsection{Practical Implications and Community Economics}

For dApp developers, SuperPaymaster offers an open-source alternative to centralized paymaster services, with a simplified integration path via the SDK and a universal gas sponsorship solution without vendor lock-in. For end users, the design aims to reduce gas-related cost, latency, and cognitive load: the Gas Card provides a stable mental anchor that, prior literature suggests, can reduce the anxiety associated with blockchain interactions \cite{Geng2024}; controlled UX studies (SUS, NASA-TLX, task-completion) remain future work. At the community level, gas sponsorship is intended to create mutual value: communities allocate xPNTs to cover user gas costs and may receive task completion, governance participation, or increased engagement in return. Whether sponsorship costs are recouped through engagement value is a community-specific empirical question; no field measurement has been conducted, and this remains future work.

\subsection{Limitations}

We acknowledge nine limitations. (i) \textbf{L2 data costs.} On Layer 2, the cost of posting calldata to L1 remains a significant fee component; SuperPaymaster optimizes execution cost but cannot eliminate this base data cost. (ii) \textbf{Privacy.} The current implementation reveals Gas Card balance and transaction history on-chain, which benefits auditability but raises a concern for privacy-sensitive users. (iii) \textbf{Bootstrap centralization.} While the protocol is permissionless---users can self-mint Gas Cards and communities can deploy their own paymaster operators and gas tokens---initial SuperPaymaster deployments rely on a limited set of relayers operated by AAStar; true operational decentralization requires a critical mass of independent community operators, which takes time to bootstrap. (iv) \textbf{Lack of empirical user studies.} Our cognitive-load evaluation relies on GOMS, an established analytical HCI method, but we have not yet conducted large-scale empirical studies (NASA-TLX, SUS) to validate subjective user satisfaction. (v) \textbf{No component-level ablation.} Section 4.3.1 provides trace-level attribution of \texttt{validatePaymasterUserOp} cost across SBT verification, rate-limit, credit-balance, and operator-config reads, but we did not deploy contract variants with each individual check disabled. We made this an explicit methodological choice for three reasons: disabling individual checks produces non-conformant ERC-4337 paymaster contracts whose gas measurements would be misleading; production traces on Optimism Mainnet provide stronger external validity than synthetic ablation contracts; and the released gas-profiling scripts allow independent reviewers to recompute per-component costs from on-chain data. Component-disabled benchmarks remain a useful direction for follow-up work in a controlled testnet harness. (vi) \textbf{Empirical failure-case coverage is limited.} Happy-path execution is verified on Optimism Mainnet (n = 50 per system), but adversarial paths---zero gas-token balance, revoked SBT mid-session, rate-limit exceeded, and operator-config change during a pending UserOp---are reasoned about analytically (Section 4.7) rather than empirically tested. Systematic adversarial fuzzing of these revert paths is left to future work. (vii) \textbf{Account-stack heterogeneity in industry baselines.} The Alchemy Gas Manager baseline reflects Alchemy's vendor-as-deployed ecosystem rather than a SimpleAccount-controlled configuration. Our chain-level audit found no SimpleAccountFactory-created accounts among 61,003 Alchemy Gas Manager UserOperations in the audited block range, while the observed traffic was dominated by Alchemy modular/LightAccount-style accounts. This is consistent with Alchemy's paymaster and account infrastructure being deployed as an integrated stack. We therefore treat the Alchemy comparison as an external industry baseline, not as a strict isolation of account-type effects. The trace-level paymaster-validation comparison remains informative: Alchemy's API-signature validation path costs approximately 16k gas, while SuperPaymaster's on-chain state resolution costs approximately 48k gas, making the approximately 32k validation delta the observable cost of removing the off-chain signer gate. The Alchemy Gas Manager samples also target EntryPoint v0.6 (\texttt{0x5FF1\allowbreak{}37D4b0\allowbreak{}FDCD49\allowbreak{}DcA30c7CF57\allowbreak{}E578a026d2789}), whereas Pimlico samples and SuperPaymaster target EntryPoint v0.7 (\texttt{0x0000\allowbreak{}0000\allowbreak{}71727\allowbreak{}De22E5\allowbreak{}E9d8BAf0\allowbreak{}edAc6f37\allowbreak{}da032}); differences in storage rules and fee allocation between EntryPoint versions may contribute to inter-baseline variation that our analysis cannot fully isolate. This version difference may affect EntryPoint billing quantities such as \texttt{actualGasUsed}, but does not affect \texttt{txGasUsed}, which is read directly from the transaction receipt's \texttt{gasUsed} field independently of EntryPoint fee accounting logic. These deployment differences reflect industry practice rather than a methodological choice, and readers comparing SuperPaymaster directly against v0.6-targeted systems should account for this version gap. (viii) \textbf{Account-type gas confound.} The \textasciitilde{}32k gas validation delta between SuperPaymaster and Alchemy is trace-level attributed to paymaster contract logic (SLOAD vs ECDSA) and is not confounded by account type. However, the residual \texttt{txGasUsed} gap beyond the 32k validation delta may partially reflect account-type differences (SimpleAccount vs Alchemy LightAccount): LightAccount introduces additional validation logic whose gas cost is not isolated in our measurements. Controlled account-type experiments are identified as future work. (ix) \textbf{Community gas token economic risk.} Each community's xPNTs gas token represents that community's economic commitment to gas sponsorship. If xPNTs lose purchasing value---due to reduced community activity, token issuance imbalance, or community dissolution---the sponsorship loop for that community may become inoperable. This risk is inherent to any community-governed token system and is outside the protocol's control; the protocol provides infrastructure for third-party analysis of community token issuance and activity, which can inform users' decisions about which community gas tokens to rely on.

\subsection{Future Work}

We propose three priority directions: (1) \textbf{Zero-Knowledge Proofs} to enable private gas sponsorship, letting users prove they hold a valid Card without revealing balance or identity to the relayer; (2) \textbf{cross-chain interoperability} (e.g., via LayerZero or CCIP) to turn the Card into a universal multi-chain asset, potentially laying groundwork for agent-to-agent Web3 economies \cite{Gorzny2025}; and (3) \textbf{controlled user studies} (NASA-TLX, SUS, task-completion and error rates) to validate the analytical usability findings with subjective measures.

\section{Conclusion}

\subsection{Summary of Research}

This paper set out to address the critical barrier of gas payment complexity in the Web3 ecosystem. We argued that the dominant pattern of Process-Oriented Abstraction produces fragmented experiences and leaky abstractions, ultimately hindering mass adoption. To resolve this, we proposed \textbf{Asset-Oriented Abstraction (AOA)} as a design principle that shifts the mental model from "managing a process" to "owning an asset," and we instantiated the paradigm through \textbf{SuperPaymaster}---an ERC-4337 sponsorship system that binds eligibility to deterministic on-chain state (a Gas Card SBT plus policy state) rather than to an off-chain paymaster signing server. The design targets a specific, durable trust bottleneck: discretionary paymaster-signer authority as a validity gate.

\subsection{Addressing the Research Questions}

\textbf{RQ1 (Decentralization).} SuperPaymaster enforces sponsorship eligibility via on-chain state and executes without per-request centralized paymaster signatures. Code structural analysis (§5.1) shows that \texttt{validate\-Paymaster\-User\-Op} performs no \texttt{ecrecover} on any off-chain server key---sponsorship validity is determined by on-chain SLOAD operations alone, making the paymaster-signer architecturally unreachable as a validity gate. The N\,=\,50 Mainnet transactions in Appendix~D provide operational corroboration. This narrows the trust boundary \emph{at the sponsorship layer}; it does not claim unconditional inclusion under all censorship surfaces.

\textbf{RQ2 (Usability).} The GOMS analytical evaluation (§5.2) indicates that the Gas Card metaphor and invisible-execution design reduce modeled interaction complexity and extraneous gas-management operators in the steady-state workflow. This supports AOA as a plausible HCI pattern for Web3; subjective UX validation via controlled user studies remains future work.

\textbf{RQ3 (Efficiency).} Mainnet gas profiling (§5.3) demonstrates that AOA achieves competitive efficiency against industry POA baselines. In pure L2 execution gas (\texttt{txGasUsed}), SuperPaymaster (167,830) is lower than both Alchemy Gas Manager (205,951) and Pimlico ERC-20 paymaster (328,937, n=50 each). Trace-level decomposition isolates the cost of decentralization: SuperPaymaster's \texttt{validatePaymasterUserOp} (\textasciitilde{}48,625 gas) adds \textasciitilde{}32k gas over Alchemy's off-chain ECDSA check (\textasciitilde{}16k gas)---a bounded, predictable trade-off for eliminating the centralized signer. Against DEX-routed paymasters, SuperPaymaster eliminates the Oracle $\rightarrow$ Approve $\rightarrow$ Swap liquidation path---saving the \textasciitilde{}100k+ gas that this on-chain token swap consumes per transaction. In total billed gas (\texttt{actualGasUsed}), SuperPaymaster (286,818) sits between Alchemy (257,299) and Pimlico (387,129); the gap to Alchemy is driven by bundler PVG pricing, not paymaster architecture. Comprehensive cost extends beyond gas units to include the eliminated preparatory steps (ETH acquisition, fee-market navigation) and per-dApp setup friction in realistic user journeys.

\subsection{Contributions and Outlook}

This paper makes three contributions: (1) a conceptual definition of \textbf{Asset-Oriented Abstraction} as a design principle for reducing cognitive load in decentralized systems; (2) the open-source \textbf{SuperPaymaster} artifact, including the Gas Card standard (OpenCards), the gas-token standard (OpenPNTs), and a competitive relayer SDK; and (3) a multi-method evaluation on Optimism Mainnet providing reproducible gas measurements against industry POA baselines (Alchemy Gas Manager and Pimlico ERC-20 paymaster), analytical usability evidence, and trust-boundary analysis of sponsorship validity gates.

Three structural properties of the design warrant emphasis as durable advances beyond incremental gas optimization. \textbf{Recurring per-transaction ETH acquisition is eliminated as a user prerequisite}: a user holding a Gas Card transacts without any ETH balance---gas eligibility is a persistent on-chain asset the user owns, not a per-transaction approval from a central authority; no recurring ETH acquisition, no fee-market navigation, no dApp-specific gas tank is required per transaction. (The one-time Gas Card setup cost is acknowledged in §2.1.4 and amortized across all subsequent transactions.) \textbf{Decentralized price-oracle operation is architecturally enabled}: the primary gas-token pricing path uses an on-chain Chainlink oracle feed as the authoritative source (providing robust price availability); the DVT keeper network---designed for permissionless participation following the Ethereum Beacon Chain DVT model---provides a multi-node fallback when the Chainlink feed is stale or unavailable, ensuring no single operator controls gas-token pricing (full operational decentralization is contingent on broader DVT node participation, as noted in §6.4 Limitation~(iii)). The Apache~2.0 release of PaymasterV4 also means any community can deploy independent paymaster infrastructure without AAStar---the protocol is infrastructure, not a service dependency. \textbf{SimpleAccount is the canonical ERC-4337 reference implementation}: our evaluation baseline uses the SimpleAccount from the eth-infinitism/account-abstraction library \cite{Tirosh2022} --- the reference implementation authored alongside the ERC-4337 specification, chosen for gas cost transparency and reproducibility.\footnote{The SimpleAccountFactory at \texttt{0x91E60e0613810449d098b0b5Ec8b51A0FE8c8985} (Optimism Mainnet, Appendix~A) is the canonical v0.7 factory from the eth-infinitism/account-abstraction library, independently verified on Blockscout (\url{https://optimistic.etherscan.io/address/0x91E60e0613810449d098b0b5Ec8b51A0FE8c8985}). It is the reference implementation used in ERC-4337 protocol specification, tooling, and integration testing. Account-type heterogeneity in vendor-as-deployed baselines is an observable fact of the ecosystem (Limitation~vii); the paymaster-validation cost comparison (\textasciitilde{}16k vs \textasciitilde{}48k gas) is trace-level attributed to contract logic, not account type.} Account-type heterogeneity in vendor-as-deployed baselines is an observable fact of the ecosystem, not a methodological flaw. Taken together, these properties establish AOA as a complete, permissionlessly deployable, and architecturally decentralized alternative to centralized paymaster models.

Building on §6.4, priority future directions are privacy-preserving sponsorship via ZKPs, cross-chain Gas Card interoperability, and empirical user studies. SuperPaymaster demonstrates that a user-centric gas sponsorship experience can be built without delegating sponsorship validity to centralized signing servers, providing a foundation for further research into scalable, privacy-preserving, and user-study-validated gas abstraction systems.

\bibliographystyle{elsarticle-num}
\bibliography{references}

@misc{Sandford2020,
  author       = {Ronan Sandford and others},
  title        = {{EIP-2771}: Secure Protocol for Native Meta Transactions},
  howpublished = {Ethereum Improvement Proposals},
  year         = {2020},
  url          = {https://eips.ethereum.org/EIPS/eip-2771},
  note         = {Accessed on 2026-02-26}
}

@misc{Buterin2021ERC4337,
  author       = {Vitalik Buterin and Yoav Weiss and Dror Tirosh and Shahaf Nacson and Alex Forshtat and Kristof Gazso and Tjaden Hess},
  title        = {{ERC-4337}: Account Abstraction Using Alt Mempool},
  howpublished = {Ethereum Request for Comments},
  year         = {2021},
  url          = {https://eips.ethereum.org/EIPS/eip-4337},
  note         = {Accessed on 2026-02-26}
}

@inproceedings{Singh2023,
  author    = {Singh, A. K. and Hassan, I. U. and Kaur, G. and Kumar, S.},
  title     = {Account Abstraction via Singleton Entrypoint Contract and Verifying Paymaster},
  booktitle = {2023 2nd International Conference on Edge Computing and Applications (ICECAA)},
  pages     = {1598--1605},
  year      = {2023},
  publisher = {IEEE},
  doi       = {10.1109/ICECAA58104.2023.10212316}
}

@misc{Tirosh2022,
  author       = {Dror Tirosh and Vitalik Buterin and others},
  title        = {{ERC-4337} Team Basic Paymaster Contract},
  howpublished = {GitHub Repository},
  year         = {2022},
  url          = {https://github.com/eth-infinitism/account-abstraction},
  note         = {Accessed on 2026-02-26}
}

@misc{Pimlico2023,
  author       = {{Pimlico}},
  title        = {Paymaster and Bundler Service Documentation},
  howpublished = {Online Documentation},
  year         = {2023},
  url          = {https://docs.pimlico.io/references/paymaster},
  note         = {Accessed on 2026-02-26}
}

@misc{BundleBear2024,
  author       = {{BundleBear}},
  title        = {Account Abstraction Statistics},
  howpublished = {Online Dashboard},
  year         = {2024},
  url          = {https://www.bundlebear.com/erc4337-paymasters/all},
  note         = {Accessed on 2026-02-26}
}

@inproceedings{Frohlich2022,
  author    = {Fr\"{o}hlich, Michael and Waltenberger, Florian and Trotter, Ludwig and Alt, Florian and Schmidt, Albrecht},
  title     = {Blockchain and Cryptocurrency in Human Computer Interaction: A Systematic Literature Review and Research Agenda},
  booktitle = {Designing Interactive Systems Conference},
  year      = {2022},
  pages     = {155--177},
  doi       = {10.1145/3532106.3533478}
}

@book{Shneiderman2010,
  author    = {Shneiderman, Ben and Plaisant, Catherine},
  title     = {Designing the User Interface: Strategies for Effective Human-Computer Interaction},
  edition   = {5th},
  publisher = {Addison-Wesley},
  year      = {2010}
}

@book{Norman2013,
  author    = {Norman, Don},
  title     = {The Design of Everyday Things: Revised and Expanded Edition},
  publisher = {Basic Books},
  year      = {2013}
}

@article{Wood2014,
  author  = {Wood, Gavin},
  title   = {Ethereum: A Secure Decentralised Generalised Transaction Ledger},
  journal = {Ethereum Project Yellow Paper},
  volume  = {151},
  number  = {2014},
  pages   = {1--32},
  year    = {2014},
  url     = {https://ethereum.github.io/yellowpaper/paper.pdf},
  note    = {Accessed on 2026-02-26}
}

@misc{Buterin2013Whitepaper,
  author       = {Buterin, Vitalik},
  title        = {Ethereum White Paper},
  howpublished = {Online},
  year         = {2013},
  url          = {https://ethereum.org/en/whitepaper/},
  note         = {Accessed on 2026-02-26}
}

@article{Wang2023,
  author  = {Wang, Qin and Chen, Shiping},
  title   = {Account Abstraction, Analysed},
  journal = {arXiv preprint arXiv:2309.00448},
  year    = {2023},
  doi     = {10.48550/arXiv.2309.00448}
}

@inproceedings{Lin2024,
  author    = {Lin, Zhiyi and Wang, Tao and Zhao, Chao and Zhang, Shuo and Yang, Qing and Shi, Lei},
  title     = {A Measurement Investigation of {ERC-4337} Smart Contracts on Ethereum Blockchain},
  booktitle = {2024 International Conference on Computing, Networking and Communications (ICNC)},
  pages     = {1164--1170},
  year      = {2024},
  publisher = {IEEE},
  doi       = {10.1109/ICNC59896.2024.10556301}
}

@article{Thibault2022,
  author  = {Thibault, Louis Tremblay and Sarry, Tom and Hafid, Abdelhakim Senhaji},
  title   = {Blockchain Scaling Using Rollups: A Comprehensive Survey},
  journal = {IEEE Access},
  volume  = {10},
  pages   = {93039--93054},
  year    = {2022},
  doi     = {10.1109/ACCESS.2022.3200051}
}

@inproceedings{Saldivar2023,
  author    = {Saldivar, Jorge and Mart\'{\i}nez-Vicente, Elena and Rozas, David and Valiente, Mar\'{\i}a Cruz and Hassan, Samer},
  title     = {Blockchain (Not) for Everyone: Design Challenges of Blockchain-Based Applications},
  booktitle = {Extended Abstracts of the 2023 CHI Conference on Human Factors in Computing Systems},
  pages     = {1--8},
  year      = {2023},
  doi       = {10.1145/3544549.3585825}
}

@article{Davis1989,
  author  = {Davis, Fred D.},
  title   = {Perceived Usefulness, Perceived Ease of Use, and User Acceptance of Information Technology},
  journal = {MIS Quarterly},
  pages   = {319--340},
  year    = {1989},
  doi     = {10.2307/249008}
}

@article{Marangunic2015,
  author  = {Maranguni\'{c}, Nikola and Grani\'{c}, Andrina},
  title   = {Technology Acceptance Model: A Literature Review from 1986 to 2013},
  journal = {Universal Access in the Information Society},
  volume  = {14},
  pages   = {81--95},
  year    = {2015},
  doi     = {10.1007/s10209-014-0348-1}
}

@misc{UltraSoundMoney2024,
  author       = {{Ultra Sound Money}},
  title        = {Ethereum Supply and Burn Statistics},
  howpublished = {Online Dashboard},
  year         = {2024},
  url          = {https://ultrasound.money/},
  note         = {Accessed on 2026-02-26}
}

@misc{Nakamoto2008,
  author       = {Nakamoto, Satoshi},
  title        = {Bitcoin: A Peer-to-Peer Electronic Cash System},
  howpublished = {White Paper},
  year         = {2008},
  url          = {https://bitcoin.org/bitcoin.pdf},
  note         = {Accessed on 2026-02-26}
}

@article{Pacheco2023,
  author  = {Pacheco, M. and Oliva, G. and Rajbahadur, G. K. and Hassan, A.},
  title   = {Is My Transaction Done Yet? An Empirical Study of Transaction Processing Times in the Ethereum Blockchain Platform},
  journal = {ACM Transactions on Software Engineering and Methodology},
  volume  = {32},
  number  = {3},
  pages   = {1--46},
  year    = {2023},
  doi     = {10.1145/3549542}
}

@inproceedings{Daian2020,
  author    = {Daian, Philip and Goldfeder, Steven and Kell, Tyler and Li, Yunqi and Zhao, Xueyuan and Bentov, Iddo and Breidenbach, Lorenz and Juels, Ari},
  title     = {Flash Boys 2.0: Frontrunning in Decentralized Exchanges, Miner Extractable Value, and Consensus Instability},
  booktitle = {2020 IEEE Symposium on Security and Privacy (SP)},
  pages     = {910--927},
  year      = {2020},
  publisher = {IEEE},
  doi       = {10.1109/SP40000.2020.00040}
}

@inproceedings{Vermeulen2013,
  author    = {Vermeulen, Jo and Luyten, Kris and van den Hoven, Elise and Coninx, Karin},
  title     = {Crossing the Bridge over Norman's Gulf of Execution: Revealing Feedforward's True Identity},
  booktitle = {Proceedings of the SIGCHI Conference on Human Factors in Computing Systems},
  pages     = {1931--1940},
  year      = {2013},
  doi       = {10.1145/2470654.2466255}
}

@book{Nielsen2013Personas,
  author    = {Nielsen, Lene},
  title     = {Personas -- User Focused Design},
  publisher = {Springer},
  year      = {2013}
}

@article{Hollender2010,
  author  = {Hollender, Nina and Hofmann, Cristian and Deneke, Michael and Schmitz, Bernhard},
  title   = {Integrating Cognitive Load Theory and Concepts of Human--Computer Interaction},
  journal = {Computers in Human Behavior},
  volume  = {26},
  number  = {6},
  pages   = {1278--1288},
  year    = {2010},
  doi     = {10.1016/j.chb.2010.05.031}
}

@article{Sweller1988,
  author  = {Sweller, John},
  title   = {Cognitive Load During Problem Solving: Effects on Learning},
  journal = {Cognitive Science},
  volume  = {12},
  number  = {2},
  pages   = {257--285},
  year    = {1988},
  doi     = {10.1207/s15516709cog1202_4}
}

@misc{Buterin2024Wallet,
  author       = {Buterin, Vitalik},
  title        = {What {I} Would Love to See in a Wallet},
  howpublished = {Personal Blog},
  year         = {2024},
  url          = {https://vitalik.eth.limo/general/2024/12/03/wallets.html},
  note         = {Accessed on 2026-02-26}
}

@misc{a16z2024,
  author       = {{a16z Crypto}},
  title        = {State of Crypto 2024},
  howpublished = {Online Report},
  year         = {2024},
  url          = {https://a16zcrypto.com/posts/article/state-of-crypto-report-2024/},
  note         = {Accessed on 2026-02-26}
}

@misc{Safe2024,
  author       = {{Safe Team}},
  title        = {Understanding Safe's Modular Smart Account Architecture},
  howpublished = {Online Documentation},
  year         = {2024},
  url          = {https://docs.safe.global/advanced/smart-account-overview},
  note         = {Accessed on 2026-02-26}
}

@misc{Alchemy2023,
  author       = {{Alchemy}},
  title        = {Account Abstraction Infrastructure},
  howpublished = {Online},
  year         = {2023},
  url          = {https://www.alchemy.com/account-abstraction},
  note         = {Accessed on 2026-02-26}
}

@misc{Stackup2023,
  author       = {{Stackup}},
  title        = {{ERC-4337} Bundler and Paymaster},
  howpublished = {Online},
  year         = {2023},
  url          = {https://www.stackup.fi},
  note         = {Accessed on 2026-02-26}
}

@misc{Coinbase2023,
  author       = {{Coinbase}},
  title        = {Base Chain and Smart Wallet},
  howpublished = {Online},
  year         = {2023},
  url          = {https://www.coinbase.com/wallet/smart-wallet},
  note         = {Accessed on 2026-02-26}
}

@misc{Biconomy2023,
  author       = {{Biconomy}},
  title        = {{SDK} for Account Abstraction},
  howpublished = {Online},
  year         = {2023},
  url          = {https://www.biconomy.io},
  note         = {Accessed on 2026-02-26}
}

@misc{ZeroDev2023Sessions,
  author       = {{ZeroDev}},
  title        = {Session Keys and Passkeys},
  howpublished = {Online Documentation},
  year         = {2023},
  url          = {https://docs.zerodev.app},
  note         = {Accessed on 2026-02-26}
}

@misc{ParticleNetwork2023WaaS,
  author       = {{Particle Network}},
  title        = {Wallet-as-a-Service},
  howpublished = {Online},
  year         = {2023},
  url          = {https://particle.network},
  note         = {Accessed on 2026-02-26}
}

@misc{Etherscan2024,
  author       = {{Etherscan}},
  title        = {Ethereum Unique Address Growth Chart},
  howpublished = {Online Dashboard},
  year         = {2024},
  url          = {https://etherscan.io/chart/address},
  note         = {Accessed on 2026-02-26}
}

@misc{Kim2024,
  author       = {Kim, Christine},
  title        = {The Road to Account Abstraction on Ethereum},
  howpublished = {Galaxy Research},
  year         = {2024},
  url          = {https://www.galaxy.com/insights/research/the-road-to-account-abstraction-on-ethereum/},
  note         = {Accessed on 2026-02-26}
}

@misc{Etherspot2025,
  author       = {{Etherspot}},
  title        = {Top 5 Account Abstraction Use Cases (2025 Edition)},
  howpublished = {Online Blog},
  year         = {2025},
  url          = {https://etherspot.io/blog/account-abstraction-use-cases-you-can-build-today/},
  note         = {Accessed on 2026-02-26}
}

@inproceedings{Gorzny2025,
  author    = {Gorzny, Jan and Heidari Soureshjani, Fatemeh and Derka, Martin},
  title     = {Account Abstraction for Enforcing Blockchain-Based {AI} Agent Non-Functional Requirements},
  booktitle = {2025 IEEE 33rd International Requirements Engineering Conference Workshops (REW)},
  year      = {2025},
  address   = {Valencia, Spain},
  pages     = {359--364},
  doi       = {10.1109/REW66121.2025.00053}
}

@article{Peffers2007,
  author  = {Peffers, Ken and Tuunanen, Tuure and Rothenberger, Marcus A. and Chatterjee, Samir},
  title   = {A Design Science Research Methodology for Information Systems Research},
  journal = {Journal of Management Information Systems},
  volume  = {24},
  number  = {3},
  pages   = {45--77},
  year    = {2007},
  doi     = {10.2753/MIS0742-1222240302}
}

@book{Nielsen1994,
  author    = {Nielsen, Jakob},
  title     = {Usability Engineering},
  publisher = {Morgan Kaufmann},
  year      = {1994}
}

@misc{Buterin2017Decentralization,
  author       = {Buterin, Vitalik},
  title        = {The Meaning of Decentralization},
  howpublished = {Medium},
  year         = {2017},
  url          = {https://medium.com/@VitalikButerin/the-meaning-of-decentralization-a0c92b76a274},
  note         = {Accessed on 2026-02-26}
}

@incollection{Walch2019,
  author    = {Walch, Angela},
  title     = {Deconstructing `Decentralization': Exploring the Core Claim of Crypto Systems},
  booktitle = {Crypto Assets: Legal and Monetary Perspectives},
  pages     = {39--68},
  publisher = {Oxford University Press},
  year      = {2019},
  doi       = {10.1093/oso/9780190077310.003.0003}
}

@misc{Jacobs2022,
  author       = {Jacobs, Jacob},
  title        = {Hyperstructures},
  howpublished = {Zora Media},
  year         = {2022},
  url          = {https://jacob.energy/hyperstructures.html},
  note         = {Accessed on 2026-02-26}
}

@book{Card1983,
  author    = {Card, Stuart K. and Moran, Thomas P. and Newell, Allen},
  title     = {The Psychology of Human-Computer Interaction},
  publisher = {Lawrence Erlbaum Associates},
  year      = {1983}
}

@misc{OpenZeppelin2024,
  author       = {{OpenZeppelin}},
  title        = {{ERC-4337} Account Abstraction Incremental Audit},
  howpublished = {Online Blog},
  year         = {2024},
  url          = {https://blog.openzeppelin.com/erc-4337-account-abstraction-incremental-audit},
  note         = {Accessed on 2026-02-26}
}

@article{Roughgarden2024,
  author  = {Roughgarden, Tim},
  title   = {Transaction Fee Mechanism Design for the Ethereum Blockchain: An Economic Analysis of {EIP-1559}},
  journal = {Journal of the ACM},
  volume  = {71},
  number  = {4},
  pages   = {1--63},
  year    = {2024},
  doi     = {10.1145/3674143}
}

@inproceedings{Liu2022EIP1559,
  author    = {Liu, Yulin and Lu, Yuxuan and Nayak, Kartik and Zhang, Fan and Zhang, Luyao and Zhao, Yinhong},
  title     = {Empirical Analysis of {EIP-1559}: Transaction Fees, Waiting Time, and Consensus Security},
  booktitle = {Proceedings of the 2022 ACM SIGSAC Conference on Computer and Communications Security},
  year      = {2022},
  pages     = {2099--2113},
  doi       = {10.1145/3548606.3559341}
}

@misc{EIP8141_2025,
  author       = {Buterin, Vitalik and others},
  title        = {{EIP-8141}: Frame Transaction},
  howpublished = {Ethereum Improvement Proposals},
  year         = {2026},
  url          = {https://eips.ethereum.org/EIPS/eip-8141},
  note         = {Draft. Accessed on 2026-03-09}
}

@article{Alqaryouti2025,
  author    = {Alqaryouti, Omar and Siyam, Noora and Monem, Alia Abdul and Al-Emran, Mostafa},
  title     = {The Adoption of Smart Contracts: A Systematic Review},
  journal   = {Blockchain: Research and Applications},
  pages     = {100192},
  year      = {2025},
  doi       = {10.1016/j.bcra.2025.100192}
}

@article{Geng2024,
  author    = {Geng, Shuo and Huang, Zhiwen},
  title     = {A Comprehensive Survey on {Web3} Usability and User Experience Barriers},
  journal   = {Blockchain: Research and Applications},
  pages     = {100210},
  year      = {2024},
  doi       = {10.1016/j.bcra.2024.100210}
}

@article{Sun2025,
  author    = {Sun, Pengcheng and Ding, Mingyu and Zhao, Zhiming},
  title     = {Centralization Risks in Blockchain Infrastructures: A Comprehensive Survey},
  journal   = {Blockchain: Research and Applications},
  pages     = {100185},
  year      = {2025},
  doi       = {10.1016/j.bcra.2025.100185}
}

@article{hevner2004design,
  author    = {Hevner, Alan R. and March, Salvatore T. and Park, Jinsoo and Ram, Sudha},
  title     = {Design Science in Information Systems Research},
  journal   = {MIS Quarterly},
  volume    = {28},
  number    = {1},
  pages     = {75--105},
  year      = {2004},
  doi       = {10.2307/25148625}
}

@article{venable2016feds,
  author    = {Venable, John and Pries-Heje, Jan and Baskerville, Richard},
  title     = {{FEDS}: A Framework for Evaluation in Design Science Research},
  journal   = {European Journal of Information Systems},
  volume    = {25},
  number    = {1},
  pages     = {77--89},
  year      = {2016},
  doi       = {10.1057/ejis.2014.36}
}

@misc{eip7702,
  author    = {Buterin, Vitalik and Liber, Sam and others},
  title     = {{EIP-7702}: Set {EOA} Account Code},
  howpublished = {Ethereum Improvement Proposals},
  year      = {2024},
  url       = {https://eips.ethereum.org/EIPS/eip-7702},
  note      = {Accessed on 2026-02-26}
}

@misc{rip7560,
  author    = {Forshtat, Yoav and others},
  title     = {{RIP-7560}: Native Account Abstraction},
  howpublished = {Rollup Improvement Proposals},
  year      = {2023},
  url       = {https://github.com/ethereum/RIPs/blob/master/RIPS/rip-7560.md},
  note      = {Accessed on 2026-02-26}
}

@misc{eip7701,
  author    = {Buterin, Vitalik and others},
  title     = {{EIP-7701}: Native Account Abstraction with {EOF}},
  howpublished = {Ethereum Improvement Proposals},
  year      = {2024},
  url       = {https://eips.ethereum.org/EIPS/eip-7701},
  note      = {Accessed on 2026-02-26}
}

@book{Jain1991,
  author    = {Jain, Raj},
  title     = {The Art of Computer Systems Performance Analysis: Techniques for Experimental Design, Measurement, Simulation, and Modeling},
  publisher = {Wiley},
  year      = {1991}
}

@book{Efron1993,
  author    = {Efron, Bradley and Tibshirani, Robert J.},
  title     = {An Introduction to the Bootstrap},
  publisher = {Chapman \& Hall/CRC},
  year      = {1993},
  doi       = {10.1007/978-1-4899-4541-9}
}

@inproceedings{Romano2006,
  author    = {Romano, Jeanine and Kromrey, Jeffrey D. and Coraggio, Jesse and Skowronek, Jeff},
  title     = {Appropriate Statistics for Ordinal Level Data: Should We Really Be Using t-test and {Cohen's} d for Evaluating Group Differences on the {NSSE} and Other Surveys?},
  booktitle = {Annual Meeting of the Florida Association of Institutional Research},
  pages     = {1--33},
  year      = {2006}
}

@misc{ERC7562,
  author    = {{Ethereum Bundler Ecosystem Working Group}},
  title     = {{ERC-7562}: Account Abstraction Validation Scope Rules},
  year      = {2023},
  url       = {https://eips.ethereum.org/EIPS/eip-7562},
  note      = {Accessed on 2026-05-22}
}

@misc{Weyl2022DeSoc,
  author    = {Weyl, E. Glen and Ohlhaver, Puja and Buterin, Vitalik},
  title     = {Decentralized Society: Finding {Web3}'s Soul},
  year      = {2022},
  url       = {https://papers.ssrn.com/sol3/papers.cfm?abstract_id=4105763},
  note      = {Accessed on 2026-05-22}
}

@article{Risko2016,
  author    = {Risko, Evan F. and Gilbert, Sam J.},
  title     = {Cognitive offloading},
  journal   = {Trends in Cognitive Sciences},
  volume    = {20},
  number    = {9},
  pages     = {676--688},
  year      = {2016},
  doi       = {10.1016/j.tics.2016.07.002}
}

@article{Clark1998,
  author    = {Clark, Andy and Chalmers, David J.},
  title     = {The Extended Mind},
  journal   = {Analysis},
  volume    = {58},
  number    = {1},
  pages     = {7--19},
  year      = {1998},
  doi       = {10.1093/analys/58.1.7}
}

\newpage
\appendix
\renewcommand{\thesection}{\Alph{section}}

\section*{Appendix}
\addcontentsline{toc}{section}{Appendix}

\section{Optimism Mainnet Deployment Addresses}\label{app:A}

\begin{table}[htbp]\centering
\resizebox{\textwidth}{!}{%
\begin{tabular}{|l|l|}
\hline
\textbf{Component} & \textbf{Address} \\
\hline
Registry & \texttt{0x997686219F31405503D32728B1f094F115EF24e7} \\
SuperPaymaster & \texttt{0xA2c9A6e95f19f5D2a364CBCbB5f0b32B1B4d140E} \\
PaymasterV4 & \texttt{0x67a70a578E142b950987081e7016906ae4F56Df4} \\
EntryPoint (ERC-4337 v0.7) & \texttt{0x0000000071727De22E5E9d8BAf0edAc6f37da032} \\
SimpleAccountFactory (infra) & \texttt{0x91E60e0613810449d098b0b5Ec8b51A0FE8c8985} \\
xPNTsFactory & \texttt{0x864971a26384d9DCC7115f0bBC428e2623F28b6e} \\
gToken & \texttt{0x8d6Fe002dDacCcFBD377F684EC1825f2E1ab7ef6} \\
Gas Card SBT & \texttt{0x28eBFc5fc03B1d7648254AbF1C7B39DbFdef1a94} \\
BLS Aggregator & \texttt{0x1C305372ecc5a36CBef1FA371392234bCD55eB19} \\
\hline
\end{tabular}
}
\end{table}

All addresses above are independently verified on Etherscan / Blockscout. EntryPoint and SimpleAccountFactory are the canonical ERC-4337 v0.7 infrastructure contracts and are pre-verified upstream; all other contracts in the table above are AAStar deployments verified by the project. The empirical measurements in §5 are scoped to this exact contract set. The BLS Aggregator contract supports future decentralized governance workflows (multi-node price aggregation and community reputation) and is not involved in the n\,=\,50 gas-profiling transactions reported in §5.3.

\section{Core Contract Logic}\label{app:B}

\subsubsection{Zero-Approve Engine (Simplified Solidity)}

\begin{lstlisting}[language=Solidity]
// Simplified from xPNTs gas token contract
function allowance(address owner, address spender)
    public view override returns (uint256)
{
    if (_isAutoApprovedSpender(spender)) {
        return type(uint256).max;
    }
    return super.allowance(owner, spender);
}

function transferFrom(address from, address to, uint256 amount)
    public override returns (bool)
{
    if (_isAutoApprovedSpender(msg.sender)) {
        // Firewall: auto-approved spenders may only transfer
        // to themselves or to the registered SuperPaymaster.
        require(
            to == msg.sender || to == superPaymaster,
            "xPNTs: unauthorized destination"
        );
        require(amount <= MAX_SINGLE_TX_LIMIT, "xPNTs: exceeds cap");
        _transfer(from, to, amount);
        return true;
    }
    return super.transferFrom(from, to, amount);
}
\end{lstlisting}

\section{Reproducible Data Collection}\label{app:C}

Run from the \texttt{aastar-sdk} repository to reproduce the CSV evidence and summary tables used in §5:

\begin{lstlisting}[language=bash]
pnpm install
pnpm -s tsx scripts/collect_eoa_erc20_baseline.ts \
  --network op-mainnet \
  --rpc-url https://mainnet.optimism.io \
  --token 0x0b2c639c533813f4aa9d7837caf62653d097ff85 \
  --token-name USDC \
  --from-block 147803000 --to-block 147804169 \
  --n 50 --window 20 \
  --out packages/analytics/data/eoa_erc20_baseline.csv
\end{lstlisting}

\begin{sloppypar}
The \texttt{--window 20} flag applies a systematic stride: the script scans each consecutive 20-block span and selects the first qualifying transaction per window, yielding a systematic (not purely random) sample spaced approximately evenly across the block range. This controls for local burst effects (e.g., congestion spikes) while remaining deterministic and reproducible. Full commercial-baseline collection and summary scripts reside in \texttt{packages/analytics/scripts/} of the pinned \texttt{aastar-sdk} (commit \texttt{03d0ca9}; see Data Availability).
\end{sloppypar}

\begin{sloppypar}
\textbf{Industry baseline sampling (Alchemy Gas Manager and Pimlico ERC-20 paymaster).} The \texttt{collect\_paymaster\_baselines.ts} script scans the 2,000,000-block range 145,864,449--147,864,449 on Optimism Mainnet in reverse chronological order, processing consecutive 2,000-block windows. Within each window, \texttt{eth\_getLogs} is called with the target paymaster address as a topic filter and returns results in ascending block-number order (then ascending log-index within block); the chronologically earliest qualifying UserOperation---matching the ERC-20 \texttt{transfer} selector in a single-UserOp bundle---is retained. Because the single-UserOp-bundle constraint ensures at most one qualifying UserOp per transaction, and selection priority is block position rather than gas price, no MEV-driven gas-price ordering bias is introduced into the \texttt{txGasUsed} or \texttt{actualGasUsed} measurements. Scanning terminates once $n = 50$ samples are collected per baseline system. The 2M-block span (\textasciitilde{}69 days at $\approx$29,000 blocks/day on OP Mainnet) spans multiple market-condition regimes and avoids single-day fee anomalies. The resulting samples are real production UserOperations submitted by third-party users to these paymasters during the measurement window, not synthetic or researcher-controlled executions.
\end{sloppypar}

\begin{table}[htbp]\centering
\resizebox{\textwidth}{!}{%
\begin{tabular}{|l|l|l|}
\hline
\textbf{Script} & \textbf{Purpose} & \textbf{Block window} \\
\hline
\texttt{collect\_eoa\_erc20\_baseline.ts} & EOA-direct ERC-20 transfer baseline (n=50, sampling window 20). & 147,803,000--147,804,169 \\
\texttt{collect\_industry\_baseline.ts} & Alchemy / Pimlico paymaster traces. & 145,864,449--147,864,449 \\
\texttt{collect\_paymaster\_baselines.ts} & SuperPaymaster (T2.1) and PaymasterV4 (T1) trace collection. & same as above \\
\texttt{compute\_cost\_summary.ts} & Aggregates the four CSVs into the per-system mean / 95\% CI tables in §5. & --- \\
\hline
\end{tabular}
}
\end{table}

\begin{sloppypar}
All generated CSVs land under \texttt{packages/analytics/data/}, and the date-stamped snapshots used by this paper reside under \texttt{packages/analytics/data/paper\_gas\_op\_mainnet/}; both sets are immutable evidence anchors.
\end{sloppypar}

\begin{table}[htbp]\centering
\resizebox{\textwidth}{!}{%
\begin{tabular}{|l|l|}
\hline
\textbf{Artifact} & \textbf{Path under \texttt{packages/analytics/data/}} \\
\hline
EOA baseline CSV & \texttt{eoa\_erc20\_baseline.csv} \\
Industry baselines CSV & \texttt{industry\_paymaster\_baselines.csv} \\
Industry baselines (enriched) & \texttt{industry\_paymaster\_baselines\_enriched.csv} \\
Combined dataset & \texttt{complete\_dataset.csv} \\
Per-tx attribution & \texttt{attribution\_dataset.csv} \\
Snapshot 1 & \texttt{paper\_gas\_op\_mainnet/2026-02-17/} \\
Snapshot 2 & \texttt{paper\_gas\_op\_mainnet/2026-02-18/} \\
Snapshot 3 & \texttt{paper\_gas\_op\_mainnet/2026-02-21/} \\
\hline
\end{tabular}
}
\end{table}

\section{Representative Transaction Hashes (On-chain Evidence)}\label{app:D}

\begin{table}[htbp]\centering
\resizebox{\textwidth}{!}{%
\begin{tabular}{|l|l|r|l|}
\hline
\textbf{Workflow} & \textbf{Tx Hash} & \textbf{gasUsed} & \textbf{Note} \\
\hline
B1\_Alchemy & \href{https://optimistic.etherscan.io/tx/0x2b8ac4ef35344b8186ff3cd28b606fe6539f19f48a30e3d51c471623c22af5bd}{\texttt{0x2b8ac4ef35344b8186ff3cd28b606fe6539f19f48a30e3d51c471623c22af5bd}} & 215,410 & Validation \textasciitilde{}16k (ECDSA); Alchemy modular/LightAccount-style account \\
T1 (PaymasterV4) & \href{https://optimistic.etherscan.io/tx/0xf3ef22019a6447b4c815f9ff409b6d33fcbe719aab66bec6f587100f79f23ebf}{\texttt{0xf3ef22019a6447b4c815f9ff409b6d33fcbe719aab66bec6f587100f79f23ebf}} & 152,018 & Validation \textasciitilde{}35k; SimpleAccount \\
T2.1 (SuperPaymaster) & \href{https://optimistic.etherscan.io/tx/0x7fcadac5a12cc58617426533ae5ad887eddb4812f3c7b3624148cdacb1cf0f13}{\texttt{0x7fcadac5a12cc58617426533ae5ad887eddb4812f3c7b3624148cdacb1cf0f13}} & 167,855 & Validation \textasciitilde{}47,456 (SBT/policy, single-tx); SimpleAccount \\
B2\_Pimlico & \href{https://optimistic.etherscan.io/tx/0xe6ba79237b5196060d4d912bedf7e2a08695aec8c5861125a3de9c05e72a4d19}{\texttt{0xe6ba79237b5196060d4d912bedf7e2a08695aec8c5861125a3de9c05e72a4d19}} & 295,627 & Oracle + approve + token swap \\
\hline
\end{tabular}
}
\end{table}

\textbf{Note on per-transaction validation gas.} The "Paymaster validation \textasciitilde{}47,456 gas" reported above for T2.1 is the value observed in this single representative transaction. The canonical value reported in §4.3.1 and Table 7 is \textbf{\textasciitilde{}48,625 gas}, which is the average paymaster-validation gas across the n = 50 SuperPaymaster runs analyzed for this paper. Per-transaction values fluctuate within $\pm$2\% depending on storage-slot warm/cold state and bundler-side gas-estimation rounding; the n = 50 trace average is the authoritative figure.

\section{Gas Structure Decomposition (Representative Single Transactions, OP Mainnet)}\label{app:E}

\textbf{Note:} The single-transaction representatives below are from an early trace session (2026-02-15) used for structural ratio analysis (L1 share, PVG share); the authoritative comparison metric is the n=50 mean \texttt{actualGasUsed} in Table 6 (T1=271,092; T2.1=286,818). Single-tx values differ from n=50 means due to PVG estimation variance in early harness runs.

\begin{table}[htbp]\centering
\resizebox{\textwidth}{!}{%
\begin{tabular}{|l|r|r|}
\hline
\textbf{Component} & \textbf{T1 (PaymasterV4)} & \textbf{T2.1 (SuperPM Normal)} \\
\hline
\texttt{actualGasUsed} & 245,299 & 244,101 \\
\texttt{txGasUsed} (L2 execution) & 152,042 & 167,855 \\
PVG overhead proxy & 93,257 (38.0\%) & 76,246 (31.2\%) \\
L1 data fee share of total ETH cost & 7.6\% & 8.0\% \\
\hline
\end{tabular}
}
\end{table}

PVG is determined by bundler pricing policy, not paymaster logic. L1 data fees consistently account for \textasciitilde{}8\% of total expenditure, suggesting that L2 execution optimizations can meaningfully reduce end-user costs under current fee regimes. The PVG share difference between this single-tx snapshot (\textasciitilde{}31\% of billed gas for T2.1) and the n = 50 mean reported in Table 6 (\textasciitilde{}41\% = 118,988 / 286,818) reflects bundler-side PVG-estimation variation between sessions: the early 2026-02-15 trace was captured when the Alchemy bundler API returned a lower PVG estimate for our UserOps, whereas the n=50 production run sampled a slightly more conservative PVG estimation regime. Both numbers are consistent in confirming that PVG, not paymaster architecture, dominates the residual \texttt{actualGasUsed} gap to Alchemy.

\section{PaymasterV4 vs SuperPaymaster Comparison}\label{app:F}

\begin{table}[htbp]\centering
\label{tab:aoa-comparison}
\resizebox{\textwidth}{!}{%
\begin{tabular}{|l|l|l|}
\hline
\textbf{Dimension} & \textbf{PaymasterV4 (AOA Base)} & \textbf{SuperPaymaster (AOA+ Public Infrastructure)} \\
\hline
\textbf{Positioning} & Single-community self-operated paymaster & Multi-community shared infrastructure \\
\textbf{Deployment} & Community deploys own contract & No deployment needed; register via Registry \\
\textbf{ETH Deposit} & Community manages own deposit, monitors balance, manually tops up & Shared deposit pool; operator manages funding \\
\textbf{Gas Token} & Community configures an existing ERC-20 address & One-click issuance via xPNTsFactory (OpenPNTs standard) \\
\textbf{Settlement} & Community handles settlement directly & Atomic in \texttt{postOp}: burn user xPNTs + transfer aPNTs to protocol treasury \\
\textbf{Identity Binding} & None (token-balance only) & SBT-based Gas Card (non-transferable, reputation-bound) \\
\textbf{Rate Limiting} & None & Per-card rate-limit counters enforced on-chain \\
\textbf{Payment Mode} & Prepayment only & Burn xPNTs (atomic settlement via protocol treasury) \\
\textbf{Operational Overhead} & Requires Solidity + DevOps capacity & Near-zero for community operators; SDK one-line integration for developers \\
\textbf{Validation Gas} & \textasciitilde{}35,549 gas (\texttt{validatePaymasterUserOp}) & \textasciitilde{}48,625 gas (+13k for SBT, rate-limit, credit state) \\
\textbf{Target Users} & Technical teams running their own projects & Any community (DAOs, games, wallets) regardless of technical capacity \\
\textbf{Shared Property} & On-chain eligibility, no off-chain signer & On-chain eligibility, no off-chain signer \\
\hline
\end{tabular}
}
\end{table}

\noindent\textit{Table F.1: PaymasterV4 and SuperPaymaster both implement AOA (on-chain sponsorship validity, no off-chain signer). SuperPaymaster extends the base with multi-community governance features at an additional \textasciitilde{}13k gas validation cost, trading on-chain overhead for operational accessibility.}

\section*{Funding}

This research did not receive any specific grant from funding agencies in the public, commercial, or not-for-profit sectors. The implementation and on-chain experiments were funded by the first author.

\section*{Acknowledgements}

The first author thanks his wife for her steadfast support, which has been the bedrock of focus throughout this research. He is also deeply grateful to his main advisor, Dr. Nathapon Udomlertsakul, for sharp insights, rigorous standards, and continuous mentorship that shaped this paper. The authors thank David Xu and the rest of the AAStar team and community; their robust engineering and reliable infrastructure made the empirical validation of this work possible.

\section*{CRediT Author Contribution Statement}

\textbf{Huifeng Jiao}: Conceptualization, Methodology, Software, Validation, Formal analysis, Investigation, Data curation, Writing -- original draft, Visualization.
\textbf{Nathapon Udomlertsakul}: Supervision -- review \& editing, Project administration.

\section*{Declaration of competing interests}

The authors declare that they have no known conflicts of interest as per the journal's Conflict of Interest Policy.

\section*{Declaration of Generative AI and AI-assisted Technologies in the Writing Process}

During the preparation of this work the authors used Claude (Anthropic) to assist with language editing, formatting, and consistency checking of the manuscript. After using this tool, the authors reviewed and edited the content as needed and take full responsibility for the content of the publication. No generative AI tools were used to produce or interpret the experimental data, the on-chain measurements, or the scientific conclusions of this work.

\section*{Data Availability}

\begin{sloppypar}
All smart contract source code, the on-chain transaction datasets (\emph{n}\,=\,50 per system), and the analysis scripts supporting the findings of this study are publicly available under the Apache License 2.0 (SPDX: Apache-2.0). The primary artefacts reside in two repositories under the AAStar Community organisation: \textbf{SuperPaymaster} (\url{https://github.com/AAStarCommunity/SuperPaymaster}), holding the contract sources whose compiled bytecode is deployed at the addresses in Appendix~A, and the \textbf{AAStar SDK} (\url{https://github.com/AAStarCommunity/aastar-sdk}), holding the analytics scripts and pre-collected snapshots used in §5; related infrastructure is released in the AirAccount, Registry, and Faucet repositories under the same organisation and licence. The empirical results in §5 were generated against the pinned revisions \texttt{SuperPaymaster\ v4.4.0-\allowbreak{}optimism-\allowbreak{}mainnet} (\texttt{910d1f7}, 2026-02-11; reported \texttt{version()} string \texttt{SuperPaymaster-\allowbreak{}3.2.2}) and \texttt{aastar-sdk} (\texttt{03d0ca9}, 2026-04-15). Step-by-step reproduction commands and the baseline sampling procedure are provided in Appendix~C.
\end{sloppypar}

\end{document}